\documentclass[a4paper,onecolumn,11pt]{quantumarticle}
\pdfoutput=1

\usepackage[numbers]{natbib}
\usepackage[utf8]{inputenc}
\usepackage[T1]{fontenc}
\usepackage[english]{babel}
\usepackage{stmaryrd,amsmath, amsthm, amssymb, amsfonts, physics, braket}
\usepackage{mathtools}
\usepackage{todonotes}
\mathtoolsset{showonlyrefs}

\usepackage{graphicx}
\usepackage{xcolor}
\usepackage{subcaption}
\usepackage{tikz}
\usepackage{pgfplots}
\pgfplotsset{compat=newest}
\usetikzlibrary{pgfplots.fillbetween}
\tikzset{snake it/.style={decorate, decoration=snake}}

\usepackage{dsfont}

\definecolor{cbblue}{HTML}{648FFF}
\definecolor{cbpurple}{HTML}{785EF0}
\definecolor{cbpink}{HTML}{DC267F}
\definecolor{cborange}{HTML}{FE6100}
\definecolor{cbyellow}{HTML}{FFB000}

\newcommand{\cba}{cbblue}
\newcommand{\cbb}{cbpurple}
\newcommand{\cbc}{cbpink}
\newcommand{\cbd}{cborange}
\newcommand{\cbe}{cbyellow}

\newcommand{\PA}[1]{{#1}}

\newtheorem{remark}{Remark}

\title{A relativistic discrete spacetime formulation of 3+1 QED}

\author{Nathana\"el Eon}
\affiliation{Aix-Marseille Université, Université de Toulon, CNRS, LIS, Marseille, France}
\email{nathanael.eon@lis-lab.fr}

\author{Giuseppe Di Molfetta}
\affiliation{Aix-Marseille Université, Université de Toulon, CNRS, LIS, Marseille, France}
\email{giuseppe.dimolfetta@lis-lab.fr}

\author{Giuseppe Magnifico}
\affiliation{Dipartimento di Fisica e Astronomia “G. Galilei”, Universita` di Padova, I-35131 Padova, Italy}
\affiliation{Padua Quantum Technologies Research Center, Universita` degli Studi di Padova}
\affiliation{Istituto Nazionale di Fisica Nucleare (INFN), Sezione di Padova, I-35131 Padova, Italy}
\affiliation{Dipartimento di Fisica, Universita` di Bari, I-70126 Bari, Italy}

\author{Pablo Arrighi}
\affiliation{Université Paris-Saclay, Inria, CNRS, LMF, 91190 Gif-sur-Yvette, France}

\begin{document}

\maketitle
\begin{abstract}
This work provides a relativistic, digital quantum simulation scheme for both $2+1$ and $3+1$ dimensional quantum electrodynamics (QED), based on a discrete spacetime formulation of theory. It takes the form of a quantum circuit, infinitely repeating across space and time, parametrised by the discretization step $\Delta_t=\Delta_x$. Strict causality \PA{at each step} is ensured as circuit wires coincide with the lightlike worldlines of QED; simulation time under decoherence is optimized. The construction replays the logic that leads to the QED Lagrangian. Namely, it starts from the Dirac quantum walk, well-known to converge towards free relativistic fermions. It then extends the quantum walk into a multi-particle sector quantum cellular automata in a way which respects the fermionic anti-commutation relations and the discrete gauge invariance symmetry. Both requirements can only be achieved at cost of introducing the gauge field. Lastly the gauge field is given its own electromagnetic dynamics, which can be formulated as a quantum walk at each plaquette. 

\end{abstract}

\section{Introduction}

The simulation of quantum phenomena, in order to scale in size, requires the use of a  quantum mechanical device. A short term application is to find ground states of Hamiltonians, that hold the key to certain molecular structures or condensed matter properties \cite{lloyd1996universal}. In the longer term, however, it may even be used to simulate the dynamics of these from first principles, based on their constituent fundamental particles' dynamics. This has motivated a strand of works on the quantum simulation of quantum field theories (QFT)  \cite{jordan2012quantum,banuls2020simulating,Preskill2019SimulatingQF}. All of them rely on a prior spatial discretization, but some are actually based on a spacetime discretization, allowing for a natively discrete account of both the relativistic and gauge symmetries. This work provides a quantum simulation scheme based on a discrete spacetime formulation of $3+1$-quantum electrodynamics. 

\textit{Lattice QFT.} Lattice gauge theories appear for instance in quantum error correction theory (e.g. Kitaev's toric code \cite{Kitaev2003FaultTQ,Savary2017QuantumSL,Gonzalez2022hardware}) and condensed matter (e.g. through their applications in spin liquids). But they originate from particle physics, where lattice quantum chromodynamics \cite{knechtli2017lattice} has extensively been used in order to obtain numerical values, to then be compared against experimental values from particle accelerators. This procedure is paradigmatic of the way new physics is discovered, making simulation take a central role. However, these techniques are computationally heavy. Finding a way to simulate lattice QFT efficiently and accurately through a quantum device would be a game changer. 

\textit{Non-relativistic, analogue simulation of QFT.} The standard ways to quantum simulate QFT are fundamentally non-relativistic, as they begin asymmetrically by discretizing space but not time, by means of a Kogut-Susskind Hamiltonian \cite{Kogut1975HamiltonianFO,Banks1976StrongCC}. Next the matter (fermions) and the gauge field (bosons) degrees of freedom are encoded as quantum systems on the simulating device, whose interactions will mimic those of the Hamiltonian \cite{banuls2020simulating}. These interactions are sometimes implemented as discrete-time unitaries, but even then these are short-time approximations of the Hamiltonian, as obtained by the Trotter formula under the non-relativistic $\Delta_t \ll \Delta_x$ assumption. This approach was recently realized experimentally using an ion trap architecture \cite{Martinez2016RealtimeDO}. Recent, classical but quantum-inspired tensor networks techniques, come to complement this standard approach \cite{Magnifico2021LatticeQE}. These use compact, approximate description of quantum states \cite{Ors2013API} such as the multiscale entanglement renormalization ansatz (MERA) \cite{Byrnes2002DensityMR,Zapp2017TensorNS}, discarding hopefully unwanted information about the states as they evolve, so that the description keeps a manageable size, whilst attempting to keep track of interesting ingredients, including entanglement. Tensor network techniques, however, mainly focus on finding low energy states and will inevitably hit a scalability and precision barrier when dealing with many-body states and their dynamics.

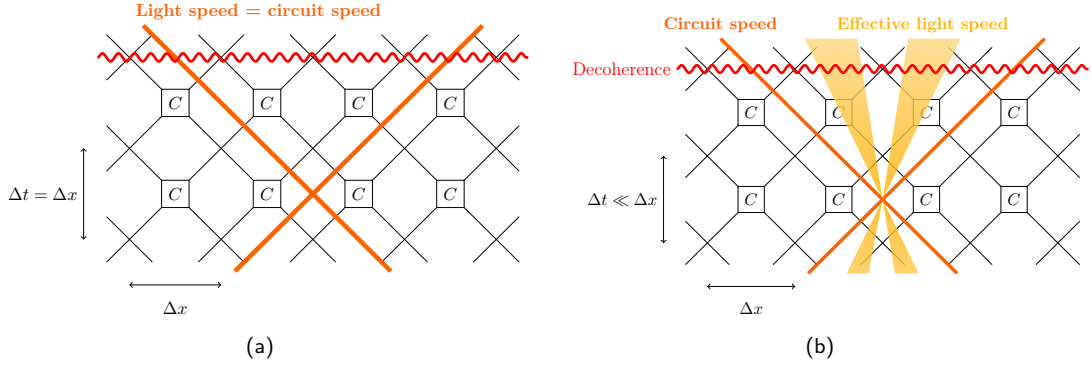
\begin{figure*}[ht!]
    \centering
    \resizebox{\dimexpr.99\textwidth}{!}{
    \begin{subfigure}{8cm}
      \centering
      \resizebox{\textwidth}{!}{\newcommand{\colora}{\cbd}
\newcommand{\colorb}{\cbb}
\newcommand{\colorc}{black}

\begin{tikzpicture}

\draw[color=black] (0,4) -- (1,5)
    (0,2) -- (3,5)
    (0,0) -- (5,5)
    (2,0) -- (7,5)
    (4,0) -- (9,5)
    (6,0) -- (9,3)
    (8,0) -- (9,1);

\draw[color=\colorc] (9,4) -- (8,5)
    (9,2) -- (6,5)
    (9,0) -- (4,5)
    (7,0) -- (2,5)
    (5,0) -- (0,5)
    (3,0) -- (0,3)
    (1,0) -- (0,1);

\draw[<->] (.5,-.5) -- (2.5,-.5);
\node at (1.5,-1) {$\Delta x$};
\draw[<->] (-.5,.5) -- (-.5,2.5);
\node[left] at (-0.5,1.5) {$\Delta t = \Delta x$};

\newcommand{\mass}[2]{
    \filldraw[color=black, fill=white] (#1+0.2,#2+0.2) rectangle (#1+0.8,#2+0.8);
    \node at (#1+0.5,#2+0.5) {$C$};
}
\foreach \i in {1,3,...,7}{
    \mass{\i}{1};
    \mass{\i}{3};
}

\draw[color=\colora, line width=1mm] 
    (2.8,-.2) -- (8.2,5.2)
    (6.2,-.2) -- (.8,5.2);
\node[color=\colora, right] at (0.5,5.5) {\textbf{Light speed = circuit speed}};

\path[draw=red, line width=.6mm, snake it] (-.2,4.5) -- (9.2,4.5);

\end{tikzpicture}}
      \caption{}
      \label{fig:sub-relativisticspacetime}
    \end{subfigure}
    \begin{subfigure}{8cm}
      \centering
      \resizebox{\textwidth}{!}{\definecolor{cboryellow}{HTML}{FE8100}
\newcommand{\colora}{\cbd}
\newcommand{\colorb}{\cbe}
\newcommand{\colorc}{black}

\begin{tikzpicture}

\draw[color=black] (0,4) -- (1,5)
    (0,2) -- (3,5)
    (0,0) -- (5,5)
    (2,0) -- (7,5)
    (4,0) -- (9,5)
    (6,0) -- (9,3)
    (8,0) -- (9,1);

\draw[color=\colorc] (9,4) -- (8,5)
    (9,2) -- (6,5)
    (9,0) -- (4,5)
    (7,0) -- (2,5)
    (5,0) -- (0,5)
    (3,0) -- (0,3)
    (1,0) -- (0,1);

\draw[<->] (.5,-.5) -- (2.5,-.5);
\node at (1.5,-1) {$\Delta x$};
\draw[<->] (-.5,.5) -- (-.5,2.5);
\node[left] at (-0.5,1.5) {$\Delta t \ll \Delta x$};

\newcommand{\mass}[2]{
    \filldraw[color=black, fill=white] (#1+0.2,#2+0.2) rectangle (#1+0.8,#2+0.8);
    \node at (#1+0.5,#2+0.5) {$C$};
}
\foreach \i in {1,3,...,7}{
    \mass{\i}{1};
    \mass{\i}{3};
}

\coordinate (o) at (4.5,1.5);

\coordinate (ca) at (.8,5.2);
\coordinate (cb) at (8.2,5.2);
\coordinate (cc) at (2.8,-.2);
\coordinate (cd) at (6.2,-.2);

\coordinate (la) at (3.32,5.2);
\coordinate (la1) at (2.72,5.2);
\coordinate (la2) at (3.92,5.2);
\coordinate (lb) at (5.68,5.2);
\coordinate (lb1) at (5.08,5.2);
\coordinate (lb2) at (6.28,5.2);

\coordinate (lc) at (3.95,-.2);
\coordinate (lc1) at (3.68,-.2);
\coordinate (lc2) at (4.18,-.2);
\coordinate (ld) at (5.05,-.2);
\coordinate (ld1) at (4.78,-.2);
\coordinate (ld2) at (5.32,-.2);

\draw[name path=A1,color=\colora, line width=.8mm] (ca) -- (cd);
\draw[name path=A2,color=\colora, line width=.8mm] (cb) -- (cc);
\node[color=\colora, above] at (ca) {\textbf{Circuit speed}};

\node[color=\colorb, above right] at (la) {\textbf{Effective light speed}};

\newcommand{\shadeopacity}{0.8}

\shade[top color=\colorb!50!white, bottom color=\colorb, opacity=\shadeopacity] (la1) -- (la2) -- (o) -- cycle;
\shade[bottom color=\colorb!70!white, top color=\colorb, opacity=\shadeopacity] (ld1) -- (ld2) -- (o) -- cycle;

\shade[top color=\colorb!50!white, bottom color=\colorb, opacity=\shadeopacity] (lb1) -- (lb2) -- (o) -- cycle;
\shade[bottom color=\colorb!70!white, top color=\colorb, opacity=\shadeopacity] (lc1) -- (lc2) -- (o) -- cycle;

\path[draw=red, line width=.6mm, snake it] (-.2,4.5) -- (9.2,4.5);
\node[color=red, left] at (-.2,4.5) {\large Decoherence};

\end{tikzpicture}}
      \caption{}
      \label{fig:sub-timetrotterizedspacetime}
    \end{subfigure}
    }
    \caption{(\subref{fig:sub-relativisticspacetime}) In relativistic, digital quantum simulation, the light-like worldlines of the simulated theory coincides with circuit wires, yielding strict causality. (\subref{fig:sub-timetrotterizedspacetime}) In non-relativistic, Trotterized analogue quantum simulation, light-like worldlines are approximately recovered through a Lieb-Robinson bound, and are slower. Thus, the simulation is running slower. As typical decoherence times match the depth of the circuit, the QFT is simulated over a shorter period.}
    \label{fig:relativisticvsnon}
\end{figure*}

\textit{Relativistic, digital simulation of QFT.} In order to quantum simulate QFT in a relativistic manner, we must place space and time on an equal footing, discretizing both simultaneously, with parameter $\Delta_t = \Delta_x $. This leads of an infinitely repeating quantum circuit, across space and time, namely a quantum cellular automata (QCA). The speed of light in the simulated QFT will then, at each step, strictly correspond with the `circuit speed', i.e. the maximal speed allowed by the wires. \PA{And whilst it is true that after multiple steps the circuit speed could potentially produce square lightcones instead of round lightcones, a robust continuum limit argument shows that this does not happen for Dirac QCA \cite{Arrighi2014TheDE}. In fact such schemes have been shown to recover discrete counterparts to Lorentz covariance in three different ways \cite{arrighi2014discrete,PaviaLORENTZ2,DebbaschLORENTZ}.
This is in stark contrast with the earlier mentioned analogue simulation paradigm, where the $\Delta_t \ll \Delta_x$ assumption kills off Lorentz covariance and} yields a non-strict and much lower speed of light for the simulated QFT, hopefully matching the Lieb-Robinson bound \cite{osborne2019continuum}---a fragile process however \cite{eisert2009supersonic,cheneau2020speed}. Fig. \ref{fig:relativisticvsnon} illustrates the circuit and light speed under both simulation paradigms. From a theoretical standpoint, relativistic, digital quantum simulation is therefore advantageous: (i) strict causality at each step is ensured as the circuit wires match the lightlike worldlines of the simulated QFT ; (ii) space and time are treated on an equal footing as demanded by relativity. 
From a practical standpoint these advantages are expected to translate into longer simulation times. For instance, where a quantum simulation device suffers from a given typical decoherence time $\tau$, we could still simulate the QFT over a period of logical time of about $\tau$.

\textit{A natively discrete approach to QFT.} In the continuous, relativistic settings, the standard way to express a QFT is by means of a Lagrangian, i.e. a `local cost function', which integrated over a possible history provides the action. \PA{The action is to be minimized in the classical theory, or to serve as phase in the quantum theory.} The use of a particular Lagrangian is justified by means of relativistic and gauge symmetries. But the way it is then brought to quantum theory is only through regularization, typically through an energy cut-off, which is essentially a discretization of spacetime. Then a continuum limit is worked out on a per-case basis through renormalization, a process which is heavily dependent on the cut-off parameters. Hence, even though the Lagrangian approach starts as continuous, it does not actually solve the problem of obtaining a well-defined quantum theory in the continuum---\PA{that is except in the case of non-interactive theories.} In general, given some Lagrangian, an elaborate continuum limiting process still needs to be worked out. Our aim is to acknowledge this fact, and instead express the QFT directly as a family of infinitely repeating circuits of local quantum gates, parametrised by the discretization step. In order to justify the use of a particular QCA we must then, just like in the Lagrangian approach, begin with a quantum walk (QW) accounting for free relativistic fermions and then extend it to the multi-particle sector QCA by imposing the fermionic anti-commutation relations as well as discrete gauge invariance, thereby deducing the need for a gauge field. We can then `turn on' the interaction by providing the gauge field with a simple dynamics. That is, we must transpose the logics of construction that leads to a particular QFT in the Lagrangian approach, to a natively discrete setting, whose discretization parameter can then be made arbitrary small. This ought to provide a rigorous, natively discrete formulation of QFT.

\textit{Closest work.} $1+1$ quantum electrodynamics (QED), also known as the Schwinger model \cite{Schwinger1962GaugeIA}, has been recovered under the relativistic, digital quantum simulation paradigm, by discretizing through $\Delta_t = \Delta_x $ and following gauge theoretical justifications in \cite{Arrighi2020AQC}. Next, this was generalized by \cite{di2020quantum,manighalam2021continuous,Sellapillay2022ADR} in order to allow for arbitrary $\Delta_t \leq \Delta_x$, so that both the continuous spacetime limits (which exists when the interaction is turned off) and the continuous time discrete space limits (which always exists) may be taken, the latter coinciding with the Kogut-Susskind Hamiltonian. The aim of the present paper is to take the work of \cite{Arrighi2020AQC} to two and three spatial dimensions, yielding the first natively discrete spacetime formulation of a `real-life' QFT, namely $3+1$ QED.

\textit{Implementing the anti-commutation of fermions.} Digital quantum simulation has been very successful at describing relativistic particles in different fields \cite{di2013quantum,di2016quantum,hatifi2019quantum}, but only a handful of works deal with interacting QFT with more than one particle \cite{ahlbrecht2012molecular,PaviaMolecular,Arrighi2020AQC,Sellapillay2022ADR}.  One of the difficulties is that in order to encode multiple fermions as qubits, one must enforce the anti-commutation of their creation/annihilation operators, e.g. through the Jordan-Wigner transformation. However, this method has all the looks of breaking locality, especially as soon as one considers more than one dimension of space. This was even formulated as a no-go result \cite{Mlodinow2020QuantumFT}, stating that any QCA implementing the fermionic anti-commutation relations in two spatial dimensions would have very high internal space dimension, as in \cite{brun2020quantum,mlodinow2021fermionic}. In the tensor network community, anti-commutation, locality and low internal space dimensions do coexist, at the cost of introducing a cut-off for the gauge field and two extra fermions per links, called rishons. Moreover, in lattice gauge theory, a solution where the fermionic degrees of freedom are replaced by bosonic ones, at the cost of introducing two new fermionic degrees of freedom, has been developed \cite{Zohar2018EliminatingFM}. A similar idea, where just the parity of the gauge field is treated as a fermion, was sketched in Farrelly's PhD thesis \cite{farrelly2017insights}. The first main contribution of this paper is to combine these ideas and formalize them in the discrete spacetime setting. We introduce no extra field, but replicate the gauge field information once for each direction. For each direction, its parity provides a rishon. Then the Jordan-Wigner transform needs only be implemented locally, at the level of each site. This does allow for a QCA of low internal space dimension, while enforcing fermionic anti-commutation. Ultimately, the reason why the no-go result \cite{Mlodinow2020QuantumFT} is circumvented is the presence of the gauge field, as well as our focus on expressing the actual dynamics in local manner---rather than the creation/annihilation operators.

\textit{Fully discrete magnetic contribution.} In QED's gauge invariant states, the fermions are the sources of the gauge field lines. In one spatial dimension there is no magnetic term, lines are confined to the unique dimension, and thus they have no dynamics \cite{Arrighi2020AQC}. But in two and three spatial dimension, the Hamiltonian has an added magnetic term, a.k.a. the plaquette term. The second main contribution of this paper is to introduce two possible discrete spacetime counterparts to the plaquette term. The first proposition works by simply integrating the plaquette term in the Fourier basis. But it requires a prior cut-off in the gauge field degrees of freedom, and allows for arbitrary changes in values within that cut-off, even in one time step. The second proposition takes the form of a local quantum walk (QW)-like evolution in the local gauge field degrees of freedom of each plaquette. It does not require a prior cut-off and ensures that gauge field values only change one step at a time. Both constructions agree in the continuum limit.

\textit{Further spin-dimensions.} \PA{In one and two spatial dimensions, the Dirac equation is a PDE on a wave function having two complex amplitudes at each site, corresponding to the presence of a particle or an antiparticle. In the multi-particle settings, and because there can be no more than one particle in a given state and site, the four occupation numbers of a site are thus: no fermion, one particle, one antiparticle, and, both a particle and an antiparticle. This can be encoded as $2$ qubits in the $2+1$ Dirac QCA. Moving on to three spatial dimensions, the Dirac equation is a PDE having four complex amplitudes at each site, corresponding to the presence of a particle spin up, a particle spin down, an antiparticle spin down or an antiparticle spin up}. In the multi-particle settings, the number of qubits per site has then to be increased from $2$ to $4$ qubits, which makes the $3+1$ QCA more involved. The third main contribution is therefore to provide a construction of the $3+1$ QED QCA in spite of this added complexity. 

The paper is organized as follows. In Sec. \ref{sec:jordanwignerext} we set the conventions and show how to enforce the fermionic anti-commutation relations while allowing for a  local definition of the gauge invariant operators that govern the dynamics of the theory. In Sec. \ref{sec:evolution2d}, we gradually derive the gauge invariant dynamics of the $2+1$ QED QCA starting from the $2+1$ Dirac QCA and adding the electric and magnetic contributions---the simpler, two spatial dimensional case makes the argument clearer. In Sec. \ref{sec:evolution3d}, we reach the $3+1$ QED QCA. Finally, we provide some perspectives.

\section{Enforcing anti-commutation, locality and gauge invariance}
\label{sec:jordanwignerext}

In QFT, fermionic particles are represented by means of operators that annihilate them or create them at position $x$. These are denoted $a_x$ and $a_x^\dagger$ respectively. Applying the annihilator on quantum states takes occupation number $\ket{1}^x$ to $\ket{0}^x$ and produces the null vector otherwise. The creator takes $\ket{0}^x$ to $\ket{1}^x$ and produces the null vector otherwise. Moreover, these operators are required to have the specific anti-commutation relations $\{a_x, a^\dagger_y\}=\delta_{x,y}$ where $\{\cdot,\cdot\}$ denotes the anti-commutator and $\delta$ the Kronecker delta.

In order to obtain a quantum numerical scheme for a QFT, to be run on a generic quantum computer or some specific-purpose quantum simulation device, we need to encode the QFT degrees of freedom as a lattice of qubits. The natural point of departure is to interpret the occupation number degrees of freedom at each site (i.e. $\ket{1}^x$ vs $\ket{0}^x$), as qubits, thereby obtaining a lattice of qubits. But enforcing the fermionic anti-commutation whilst remaining local is non-trivial. 

In Subsec. \ref{sec:operators}, the impact of the fermionic anti-commutation relations upon the local operators that are needed to express the discrete time dynamics, will be carefully worked out. In order to obtain them, however, we crucially rely on there being a gauge field, as demanded by gauge invariance.

\subsection{Introducing the gauge field}
\label{sec:introducegaugefield}

The QED Lagrangian is built by considering the Dirac Lagrangian for free fermions, and then demanding that it be gauge invariant, under $U(1)$ gauge-transformations. This is impossible without introducing a new field, the gauge field, which in the case of QED turns out to be the electromagnetic field. 

We will proceed in the same manner in the discrete. One reason for that is that in numerical analysis, the fact that a numerical scheme conserves the original symmetries is desirable and seen as a good sign of numerical stability. The other reason is more fundamental, as we aim to show that a natively discrete spacetime formulation of QED is just as legitimate at the Lagrangian formulation, in terms of its justification through symmetries. 

Discrete gauge invariance in the context of classical cellular automata has been formalized in \cite{arrighi2018gauge,Arrighi2022GaugeinvarianceIC,ArrighiGaugeUniversality}. 
Together with the treatment of gauge invariance upon quantum walks \cite{MolfettaGaugeQW} and in lattice gauge theories in general \cite{banuls2020simulating}, it inspired \cite{Arrighi2020AQC} to formulate discrete gauge transformations in the context of QED QCA in the following manner.  

\PA{Let $\varphi:\mathbb{Z}\rightarrow \mathbb{R}$. We call discrete gauge transformation the global operator $g_{\varphi}=\bigotimes_x g(\varphi(x))$ which associates, to each position $x$, an on-site gauge transformation $g(\varphi(x))$ parametrized by $\varphi(x)$ and defined as}
\begin{align}\label{eq:qcagaugetrans}
    g(\varphi(x)) : &\ket{l}^x \mapsto e^{il\varphi(x)}\ket{l}^x \\
                    &\ket{l}^y \mapsto \ket{l}^y  \quad \quad \text{if}\ x\neq y.
\end{align}
Thus, a discrete gauge transformation $g_{\varphi}$ is essentially a space-dependent phase, exempt of any regularity requirement, applied at every point of the lattice in accordance to the occupation number at that point. To be gauge invariant, the evolution of a QCA must commute with every possible gauge transformation. 

The Dirac QCA, which solely describes moving fermions, is not gauge invariant unless we introduce the gauge field. The argument boils down to the elementary fact that, as a particle moves from position $x$ to the adjacent position $x+\eta$, 
$$\ket{1}^x\otimes\ket{0}^{x+\eta}\longmapsto \ket{0}^x\otimes\ket{1}^{x+\eta}$$
the discrete gauge transformation will trigger a phase $\varphi(x)$ applied beforehand, or a phase $\varphi(x+\eta)$ applied afterwards. It follows that fermionic transport does not commute with gauge transformations.

In order to fix this, one needs to introduce a gauge field on the links. Consider a link between sites $x$ and $x+\eta$. A gauge field is much like a doorman/bouncer counter standing on the link. It adds one whenever a particle crosses the link. Actually, in the present paper, we will not place just one bouncer per link, but two---\PA{one at each end, where the ends are denoted $x:\eta$ and $x+\eta:-\eta$ respectively}. The bouncer at $x:\eta$ counts positively the number of fermions leaving $x$ towards the link, and negatively those entering $x$ from the link. The bouncer at $(x+\eta):-\eta$ counts negatively the number entering $x+\eta$ from the link, and positively those leaving $x+\eta$ towards the link. Now fermionic transport acts as
$$\ket{1}^x\ket{l}^{x:\eta}\otimes\ket{-l}^{(x+\eta):-\eta}\ket{0}^{x+\eta}\longmapsto \ket{0}^x\ket{l+1}^{x:\eta}\otimes\ket{-l-1}^{(x+\eta):-\eta}\ket{1}^{x+\eta}$$
and the discrete gauge transformation triggers a phase $(l+1)\varphi(x)-l\varphi(x+\eta)$ regardless of whether it is applied before of after the move. 

The restriction that the two gauge fields of a link be opposite of signs, as in $\ket{l}^{x:\eta}$ and $\ket{-l}^{(x+\eta):-\eta}$, is quite natural if we think of them as holding the total number of fermions that went through the link, ever, and nothing else. Still, this restriction is not a necessity for gauge invariance. \PA{However, lifting it would allow for unnecessary degrees of freedom, that are not demanded by gauge invariance and seem to be unphysical in practice. So, we must not just ask for gauge invariance, but for `minimal' gauge invariance, i.e. the demand the gauge field be obtained as a relative gauge extension \cite{ArrighiGaugeUniversality}, making this restriction a necessity.} We therefore impose this restriction across the grid, except of course at the boundaries, where it becomes vacuous.

Placing two gauge fields per link is non-standard but has several advantages: (i) each gauge field is well-localised on a site, (ii) it gauge transforms in the same way as the fermions on that site, (iii) this is number conserving, (iv) it will help to implement the fermionic anti-commutation relations in a local manner as we will now see. 

\subsection{(Anti-)commuting annihilation and lowering operators}\label{sec:operators}

Consider the lattice generated by unit vectors describing the space directions. These vectors are denoted $\mu$, $\nu$ and $\kappa$, in two spatial dimensions, only $\mu$ and $\nu$ are used. 

At each lattice site $x$ lies a group of $d$ qubits, each stating whether a fermion in mode $j\in 0\ldots (d-1)$ is present at the site. The possible modes correspond to the number of internal degrees of freedom (e.g. the spin) of the fermions. This encoding captures the Pauli exclusion principle as there cannot be two fermions in the same mode at the same site. 

Each link between $x$ and $x+\eta$ (where $\eta\in\{\pm\mu, \pm \nu, \pm \kappa\}$) has a gauge field attached at both ends: one at $x:\eta$ and the other a $(x+\eta):-\eta$. Each gauge field lives in the Hilbert space of integers $\mathcal{H}_\mathbb{Z}$. 

The local electric counting operator, denoted $E_{x:\eta}$, is the local observable of the gauge fields, i.e. it acts as 
\begin{align}
E_{x:\eta}\ket{l}^{x:\eta} = l\ket{l}^{x:\eta} \label{eq:electric_op}
\end{align} 
and as the identity elsewhere. Following the restriction that the two gauge fields of a link be of opposite signs, one has $E_{x:\eta} = -E_{x+\eta:-\eta}$.

The lowering operator of the gauge fields is $r_{x:\eta}$. It acts as $r_{x:\eta}\ket{l}^{x:\eta} = \ket{l-1}^{x:\eta}$ and as the identity elsewhere. Most often we need to act on both the gauge fields of a link with
\begin{align}
U_{x:\eta} = r_{x:\eta}r^\dagger_{x+\eta:-\eta}.\label{eq:gaugelinklowering_op}
\end{align} 
 One has $U_{x:\eta} = U^\dagger_{x+\eta:-\eta}$---i.e. lowering the gauge field value attached to one site corresponds to raising on the other site.

\PA{The local parity observable, denoted $Z_x$, is defined by acting with the on-site operator $Z$ at site $x$, and as the identity elsewhere. On each qubit at the site, $Z$ acts like the third Pauli matrix $\sigma_3$. On each gauge field at the site, it acts as $Z\ket{2l}^{x:\eta} = \ket{2l}^{x:\eta}$ and $Z\ket{2l+1}^{x:\eta} = -\ket{2l+1}^{x:\eta}$.} 

\PA{All the operators described so far were `local' operators, by which we mean that they were of the form $L_x=M_x\otimes I$ with $M_x$ acting just at site $x$ (and sometimes direct neighbours of $x$).} We will now represent the fermionic annihilation $a_{x,j}$ and gauge lowering operators $V_{x:\eta}$ of the QFT as products of these local operators.  These implementations will not be local, as we must meet the desired (anti-)commutation relations:
\begin{align}
    \{a_{x,j},a^\dagger_{y,k}\} &= \delta_{x,y}\delta_{j,k}\label{eq:fermioncommutation}\\
    [V_{x:\eta},V_{y:\zeta}] &= 0\label{eq:bosoncommutation}\\
    [V_{x:\eta},a_{y,j}] &= 0\label{eq:bosonfermioncommutation}
\end{align}
with $\{\cdot, \cdot\}$ the anti-commutator, $[\cdot, \cdot]$ 
the commutator and $\delta_{x,y}$ corresponding to Kronecker delta.

These commutation relations are commonly implemented by means of the Jordan-Wigner (JW) transform. This will be the basis for the redefined version of the operators. However, in $2+1$ and $3+1$ dimensions, it leads to non-locality of the operators expressing dynamics if used as is \cite{Mlodinow2020QuantumFT}. In order to fix this we use the idea hinted in \cite{farrelly2017insights} and treat the parity of the gauge fields as a fermion. Moreover, we use two gauge fields per link, so that this parity plays the role of the rishons of lattice gauge theories.

\PA{Let $\prec$ denote a so-called JW order between all the fermionic occupation number qubits and gauge fields present in the model. The sole requirement to make this work is that the degrees of freedom at any given site be contiguous in the JW order, i.e. that they follow each other.} In what follows we will use, quite arbitrarily, that for a given site $x$, the fermionic modes and the gauge fields are ordered as $(x,0)\prec \ldots\prec (x,d-1) \prec x:-\mu \prec  x:\mu \prec x:-\nu \prec x:\nu \prec x:-\kappa \prec  x:\kappa$. 

We define fermionic creation and gauge lowering operators based on this order. Let $x$ be a position, $j$ a mode and $\eta$ a direction:
\begin{align}
    a_{x,j}^\dagger &= \ket{1}^{x,j}\bra{0} \prod_{y\prec (x,j)} Z_y \label{eq:fermionoperators} \\
    s_{x:\eta} &= r_{x:\eta}\prod_{y\prec x:\eta} Z_y \label{eq:gaugeoperators}
\end{align}
\begin{align}
    V_{x:\eta} =& s_{x:\eta}s^\dagger_{x+\eta: -\eta}\\
    =& r_{x:\eta} \left(\prod_{y\in \llbracket x:\eta, x+\eta:-\eta\llbracket } Z_y\right) r^\dagger_{x+\eta:-\eta}
\end{align}
where $\llbracket a, b\llbracket $ is short for $[ \min\{a,b\}, \max\{a,b\} [ $.
When $x:\eta \prec x+\eta:-\eta$, this interval does not contain $x+\eta:-\eta$ and $r_{x+\eta:-\eta}$ commutes the $Z_y$s leading to the simplification:
\begin{equation}
    V_{x:\eta} = U_{x:\eta} \prod_{y\in \llbracket x:\eta, x+\eta:-\eta\llbracket } Z_y\label{eq:vasuz}
\end{equation}

The following fermionic anti-commutation relations are ensured by the JW transform:
\begin{align}
    \{ a_{x,j} , a_{y,k}^\dagger\} &= \delta_{x,y}\delta_{j,k} \\
    \{ s_{x:\eta} , s_{y:\zeta}^\dagger\} &= \delta_{x,y} \delta_{\eta,\zeta}\label{eq:rrcommutation}\\
    \{ a_{x,j} , s_{y:\eta}^\dagger\} &= 0.\label{eq:arcommutation}
\end{align}

Since $V_{x:\eta}$ is composed of two fermion-like operators, Eqs. \eqref{eq:rrcommutation} and \eqref{eq:arcommutation} yield:
\begin{align}
    [ V_{x:\eta}, V_{y:\zeta}] &= 0 \\
    [ V_{x:\eta}, a_y] &= 0.
\end{align}
Hence, the (anti-)commutation of Eqs. \eqref{eq:fermioncommutation}, \eqref{eq:bosoncommutation} and \eqref{eq:bosonfermioncommutation} are ensured.

\subsection{Local evolution operators}\label{sec:locality}

The annihilator operators $a_{x,j}$ are not local since they have an infinite trail of $Z$ on-site operators. The lowering operators $V_{x:\eta}$ can also be non-local as soon as both ends of the link are far apart in the JW order. However, these operators are never used by themselves in the evolution. They may perhaps be used by themselves for some initial state preparation. But Physics does not require that initial state preparation be local. Moreover, in the context of initial state preparation, $a_{x,j}^\dagger$ could be interpreted as a one-step implementation of having the particle come from infinity (or from the lattice border) to its current position through a product of local steps.

What matters physically is the evolution be local. One can then consider the operators allowing to express that: (i) the movement of fermions be that of free fermionic QCA, (ii) the gauge field induce an interaction between fermions---through the electric contribution, (iii) the gauge field vibrates---through the magnetic contribution. From these requirements, we can define the simplest local gauge invariant operators (checking for gauge invariance is postponed till Subsec. \ref{sec:gaugeinv}):
\begin{align}
    &a_{x,j}^\dagger a_{x,k} \label{eq:massterm}\\
    &a_{x+\eta,j}^\dagger V_{x:\eta}^\dagger a_{x,k} \label{eq:hoppingterm}\\
    &E_{x:\eta}^2\label{eq:electricterm}\\
    &P_{x:\eta,\zeta} = V_{x:\eta} V_{x+\eta: \zeta} V_{x+\zeta: \eta}^\dagger V_{x: \zeta}^\dagger \label{eq:plaquetteterm}
\end{align}
with $\eta,\zeta \in \{\mu,\nu,\kappa\}$ and $\eta\neq \zeta$. Eq. \eqref{eq:massterm} represents a mass term (i.e. a fermionic only operation, local to a site, changing the mode), Eq. \eqref{eq:hoppingterm} corresponds to a fermion hopping term (i.e. a transport), Eq. \eqref{eq:electricterm} is the squared electric operator defined previously,  and Eq. \eqref{eq:plaquetteterm} is a plaquette term (i.e. a local vibration of the gauge field). The QED QCA evolutions will be defined based upon these local operators. 

The mass term is indeed local because it involves a pair of fermionic operators acting on the same site, cancelling out the trail of $Z$ outside this site. The electric term is local by definition.

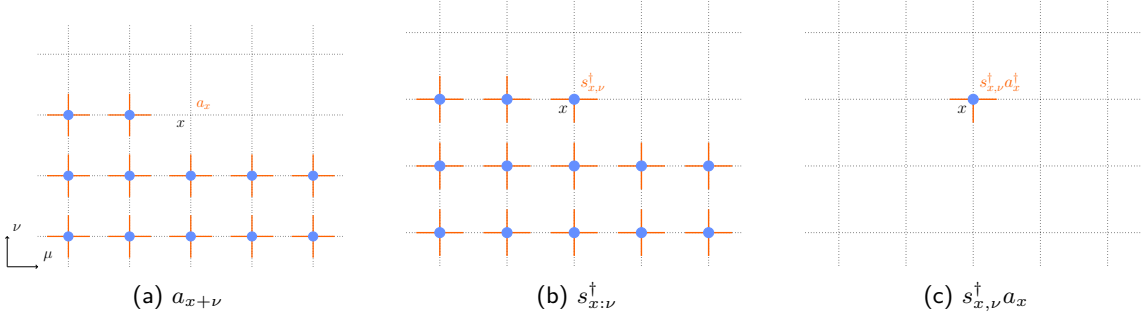
\begin{figure*}[ht!]
    \centering
    \begin{subfigure}{.3\linewidth}
      \centering
      \Large\resizebox{\linewidth}{!}{\begin{tikzpicture}

\newcommand{\affected}[2]{
    \draw[color=\cbd, very thick] (#1-0.7,#2) -- (#1,#2) -- (#1,#2-0.7);    
    \draw[color=\cbd, very thick] (#1+0.7,#2) -- (#1,#2) -- (#1,#2+0.7);   
    \filldraw[color=\cba, fill=\cba] (#1,#2) circle (0.15); 
}
\newcommand{\state}[2]{
    \filldraw[color=\cbd, fill=\cbd] (#1,#2) circle (0.15);
}

\draw[step=2.0,dotted,black,thin] (-1,-1) grid (9,7);
\affected{0}{0}
\affected{2}{0}
\affected{4}{0}
\affected{6}{0}
\affected{8}{0}
\affected{0}{2}
\affected{2}{2}
\affected{4}{2}
\affected{6}{2}
\affected{8}{2}
\affected{0}{4}
\affected{2}{4}

\node[below left,inner sep=5pt]  at (4,4) {$x$};
\node[above right,inner sep=5pt,color=\cbd]  at (4,4) {$a_{x}$};

\begin{scope}    
    \draw[->] (-2,-1) -- (-2,0);
    \draw[->] (-2,-1) -- (-1,-1);
    \node[above right,inner sep=5pt] at (-2,0) {$\nu$};
    \node[above right,inner sep=5pt] at (-1,-1) {$\mu$};
\end{scope}

\end{tikzpicture}}
      \caption{$a_{x+\nu}$}
      \label{fig:sub-localityaxeta}
    \end{subfigure}
    \hfill
    \begin{subfigure}{.3\linewidth}
      \centering
      \Large\resizebox{\linewidth}{!}{\begin{tikzpicture}

    \newcommand{\affected}[2]{
        \draw[color=\cbd, very thick] (#1-0.7,#2) -- (#1,#2) -- (#1,#2-0.7);    
        \draw[color=\cbd, very thick] (#1+0.7,#2) -- (#1,#2) -- (#1,#2+0.7);   
        \filldraw[color=\cba, fill=\cba] (#1,#2) circle (0.15); 
    }
    \newcommand{\state}[2]{
        \filldraw[color=\cbd, fill=\cbd] (#1,#2) circle (0.15);
    }

    \draw[step=2.0,dotted,black,thin] (-1,-1) grid (9,7);
    \draw[color=\cbd, very thick] (3.3,4) -- (4.7,4);
    \draw[color=\cbd, very thick] (4,4) -- (4,3.3);
    \filldraw[color=\cba, fill=\cba] (4,4) circle (0.15);
    \affected{0}{0}
    \affected{2}{0}
    \affected{4}{0}
    \affected{6}{0}
    \affected{8}{0}
    \affected{0}{2}
    \affected{2}{2}
    \affected{4}{2}
    \affected{6}{2}
    \affected{8}{2}
    \affected{0}{4}
    \affected{2}{4}

    \node[below left,inner sep=5pt]  at (4,4) {$x$};
    \node[above right,inner sep=5pt,color=\cbd]  at (4,4) {$s^\dagger_{x,\nu}$};

\end{tikzpicture}}
      \caption{$s^\dagger_{x:\nu}$}      
      \label{fig:sub-localityvxeta}
    \end{subfigure}
    \hfill
    \begin{subfigure}{.3\linewidth}
      \centering
      \Large\resizebox{\linewidth}{!}{\begin{tikzpicture}

    \newcommand{\affected}[2]{
        \draw[color=\cbd, very thick] (#1-0.7,#2) -- (#1,#2) -- (#1,#2-0.7);    
        \draw[color=\cbd, very thick] (#1+0.7,#2) -- (#1,#2) -- (#1,#2+0.7);   
        \filldraw[color=\cba, fill=\cba] (#1,#2) circle (0.15); 
    }
    \newcommand{\state}[2]{
        \filldraw[color=\cbd, fill=\cbd] (#1,#2) circle (0.15);
    }

    \draw[step=2.0,dotted,black,thin] (-1,-1) grid (9,7);
    \draw[color=\cbd, very thick] (3.3,4) -- (4.7,4);
    \draw[color=\cbd, very thick] (4,4) -- (4,3.3);
    \filldraw[color=\cba, fill=\cba] (4,4) circle (0.15);
    \node[below left,inner sep=5pt]  at (4,4) {$x$};
    \node[above right,inner sep=5pt,color=\cbd]  at (4,4) {$s^\dagger_{x,\nu}a^\dagger_{x}$};

\end{tikzpicture}}
      \caption{$s_{x,\nu}^\dagger a_{x}$}
      \label{fig:sub-localityax}
    \end{subfigure}
    \caption{Visualization of locality for $s^\dagger_{x:\nu} a_x$. The coloured dots and lines corresponds to $Z$ operators acting on fermions and gauge fields respectively. Each operator $Z_y$ is applied exactly twice which cancels them out except on site $x$. Here, only one fermionic mode per site is represented, for clarity.}
    \label{fig:locality}
\end{figure*}

For the locality of the hopping term, developing the gauge field operators $V_{x:\eta}$ as pairs $s_{x:\eta}$ and $s_{x+\eta:-\eta}^\dagger$ on each side of the link, allows for the pairing of an annihilator $a$ with an $s$ on each site, thus cancelling out the $Z$ outside the sites:
\begin{align}
    a_{x+\eta,j}^\dagger V_{x:\eta}^\dagger a_{x,k} &= a_{x+\eta,j}^\dagger (s_{x+\eta:-\eta}s_{x,\eta}^\dagger) a_{x,k} \\
        &= (a_{x+\eta,j}^\dagger s_{x+\eta:-\eta})(s_{x:\eta}^\dagger a_{x,k}) \\
        =& \ket{1}^{x+\eta,j}\bra{0} \left(\prod_{y\in \llbracket  (x+\eta,j), x+\eta:-\eta \llbracket } Z_y \right) r_{x+\eta:-\eta} r_{x:\eta}^\dagger \left(\prod_{y\in \llbracket  (x,k), x:\eta \llbracket } Z_y \right) \ket{0}^{x,k}\bra{1}\\
        =& \ket{1}^{x+\eta,j}\bra{0} \left(\prod_{y\in \llbracket  (x+\eta,j), x+\eta: -\eta \llbracket } Z_y \right) U_{x:\eta}^\dagger \left(\prod_{y\in \llbracket  (x,k), x:\eta \llbracket } Z_y \right) \ket{0}^{x,k}\bra{1}.
        \label{eq:localhoppingterm}
\end{align}
The remaining $Zs$ are local to sites $x$ and $x+\eta$. The locality of the pair $s_{x:\eta}^\dagger a_{x,k}$, which is the local brick from which the hopping term is built, is illustrated in Fig. \ref{fig:locality}. 

For the locality of plaquette term---i.e. Eq. \eqref{eq:plaquetteterm}---notice that it forms a small loop of four gauge field links. Developing each gauge operator $V_{x:\eta}$ as a pair $s_{x:\eta}$ and $s_{x+\eta:-\eta}^\dagger$, and reordering the resulting product of $s$, one gets two anti-commuting operators per site. Hence, the string of $Z$ cancels out outside the site they act on: 
\begin{align}
    P_{x:\eta,\zeta} &= V_{x:\eta} V_{x+\eta: \zeta} V_{x+\zeta: \eta}^\dagger V_{x: \zeta}^\dagger \\
    &=  \left(s_{x:\eta}s_{x+\eta:-\eta}^\dagger\right) 
        \left(s_{x+\eta:\zeta}s_{x+\eta+\zeta:-\zeta}^\dagger\right)
        \left(s_{x+\eta+\zeta:-\eta}s_{x+\zeta:\eta}^\dagger\right)
        \left(s_{x+\zeta:-\zeta}s_{x:\zeta}^\dagger\right)\\
    &=  - s_{x+\eta:-\eta}^\dagger
        \left(s_{x+\eta:\zeta}s_{x+\eta+\zeta:-\zeta}^\dagger\right)
        \left(s_{x+\eta+\zeta:-\eta}s_{x+\zeta:\eta}^\dagger\right) 
        \left(s_{x+\zeta:-\zeta} s_{x:\zeta}^\dagger\right) s_{x:\eta}\\
    &=  -\left(s_{x+\eta:-\eta}^\dagger s_{x+\eta:\zeta}\right)
        \left(s_{x+\eta+\zeta:-\zeta}^\dagger s_{x+\eta+\zeta:-\eta} \right)
        \left(s_{x+\zeta:\eta}^\dagger s_{x+\zeta:-\zeta}\right) 
        \left(s_{x:\zeta}^\dagger s_{x:\eta}\right)
\end{align}
where the minus sign that appears between the second and third line comes from the anti-commutation of $s_{x:\eta}$ with the other $7$ operators. In order to better understand the structure of the local, minus signs in this term, it is helpful to break down the plaquette term into constituent, smaller local operators. Indeed,  let us define the corner operators $c_{x:\eta,\zeta}$ as 
\begin{align}\label{eq:cornerop}
    c_{x:\eta,\zeta} &= s_{x:\zeta}^\dagger s_{x:\eta} \\
        &= r_{x:\zeta}^\dagger \left(\prod_{y\in \llbracket  x:\eta, x:\zeta \llbracket }Z_y\right) r_{x:\eta}.
\end{align}
These are local to $x$ since the remainder $Z$ is on-site $x$. A corner operator would be the result of a fermion having come from direction $\eta$, passed through site $x$, and left in the direction $\zeta$. The plaquette operator can then be redefined as:
\begin{align}\label{eq:gaugelocality}
    P_{x:\eta,\zeta} = - c_{x+\eta:\zeta,-\eta}\ c_{x+\eta+\zeta:-\eta,-\zeta}\ c_{x+\zeta: -\zeta, \eta}\ c_{x:\eta,\zeta}.
\end{align}
Thus, Eqs. \eqref{eq:localhoppingterm} and \eqref{eq:gaugelocality} allow the fermionic and bosonic dynamics terms (left-hand-side of the equations) to be expressed into simpler local dynamics (right-hand-side).

Notice that the only dependency of the definition of these local evolution operators w.r.t the JW order, is per-site. In fact the per-site JW order could even have been chosen different from one site to the other. 

\subsection{Gauge invariance}
\label{sec:gaugeinv}

Gauge invariance was introduced in Sec. \ref{sec:introducegaugefield} to motivate the need for a gauge field.
It remains to be checked that above-defined local evolution operators are gauge invariant.

\subsubsection{Gauge invariant operators}\label{sec:gioperators}

Gauge invariance is the commutation with gauge transformations, i.e. space-dependent phases proportional to the occupation number. Because the annihilation and lowering operators act on the occupation number, they individually are not gauge invariant. However, the local evolution operators combine multiple annihilation and lowering operators so that the total occupation number of a site is unchanged (only the distribution inside the site is modified)---e.g. a fermion leaving site $x$ will reduce the number of fermions by one, but increase one of the gauge fields at $x$ by one.

In order to formally show this, let us first write the commutation relations for the individual annihilation operators $a_{x,j}$
\begin{align}
    a_{x,j} g_{\varphi} &= e^{i\varphi(x)} g_{\varphi} a_{x,j} \\
    a_{x,j}^\dagger g_{\varphi} &= e^{-i\varphi(x)} g_{\varphi} a^\dagger_{x,j},
\end{align}
the gauge field operators on half-links $s_{x,\eta}$
\begin{align}
    s_{x:\eta} g_{\varphi} &= e^{i\varphi(x)} g_{\varphi} s_{x:\eta} \\
    s_{x:\eta}^\dagger g_{\varphi} &= e^{-i\varphi(x)} g_{\varphi} s_{x:\eta}^\dagger,
\end{align}
and the gauge field operators $V_{x:\eta}$
\begin{align}
    V_{x:\eta} g_{\varphi} &= e^{i(\varphi(x)-\varphi(x+\eta))} g_{\varphi} V_{x:\eta}\\
    V_{x:\eta}^\dagger g_{\varphi} &= e^{i(-\varphi(x)+\varphi(x+\eta))} g_{\varphi} V_{x:\eta}^\dagger.
\end{align}

Based on these commutation relations, we derive those of the local evolution operators. For the mass term:
\begin{align}
    \left(a_{x,j}^\dagger a_{x,k}\right) g_{\varphi} 
        &= a_{x,j}^\dagger \left( a_{x,k} g_{\varphi} \right)\\
        &= a_{x,j}^\dagger \left(e^{i\varphi(x)} g_{\varphi} a_{x,k} \right)\\
        &= e^{i\varphi(x)} \left( a_{x,j}^\dagger  g_{\varphi}\right) a_{x,k} \\
        &= e^{i\varphi(x)} \left( e^{-i\varphi(x)} g_{\varphi} a_{x,j}^\dagger \right)  a_{x,k} \\
        &= g_{\varphi} \left( a_{x,j}^\dagger a_{x,k} \right).
\end{align}
Hence, the mass term is gauge invariant.

For the hopping term:
\begin{align}
    \left(a_{x+\eta,j}^\dagger V_{x:\eta}^\dagger a_{x,j}\right) g_\varphi 
        &= a_{x+\eta,j}^\dagger V_{x:\eta}^\dagger \left( a_{x,j} g_\varphi \right) \\
        &=  a_{x+\eta,j}^\dagger V_{x:\eta}^\dagger \left(e^{i\varphi(x)} g_\varphi a_{x,j}  \right) \\
        &=  a_{x+\eta,j}^\dagger \ e^{i\varphi(x)} \left( V_{x:\eta}^\dagger  g_\varphi \right) a_{x,j}   \\
        &=  a_{x+\eta,j}^\dagger \ e^{i\varphi(x)} \left(e^{i(-\varphi(x)+\varphi(x+\eta))} g_\varphi V_{x:\eta}^\dagger  \right) a_{x,j}   \\
        &= e^{i\varphi(x+\eta)} \left(a_{x+\eta,j}^\dagger  g_\varphi \right) V_{x:\eta}^\dagger a_{x,j}   \\
        &= e^{i\varphi(x+\eta)} \left( e^{-i\varphi(x+\eta)}g_\varphi a_{x+\eta,j}^\dagger \right) V_{x:\eta}^\dagger a_{x,j}   \\
        &= g_\varphi \left( a_{x+\eta,j}^\dagger V_{x:\eta}^\dagger a_{x,j}  \right).
\end{align}
Hence, the hopping term is gauge invariant.

The electric term does not change the occupation number, hence it is directly gauge invariant:
\begin{align}
    E_{x:\eta} g_\varphi = g_\varphi E_{x:\eta}. 
\end{align}

As for the plaquette term, we start by showing the gauge invariance of the corner operator:
\begin{align}
    c_{x:\eta,\zeta} g_\varphi 
        &= s_{x:\zeta}^\dagger s_{x:\eta} g_\varphi \\
        &= s_{x:\zeta}^\dagger \left(e^{i\varphi(x)} g_\varphi s_{x:\eta}  \right) \\
        &=e^{i\varphi(x)} \left( s_{x:\zeta}^\dagger g_\varphi\right)  s_{x:\eta}  \\
        &=e^{i\varphi(x)} \left(e^{-i\varphi(x)} g_\varphi s_{x:\zeta}^\dagger \right)  s_{x:\eta}  \\
        &=  g_\varphi  \left(s_{x:\zeta}^\dagger s_{x:\eta}\right)\\
        &= g_\varphi c_{x:\eta,\zeta}.
\end{align}
Since the plaquette term is a product of four corner operators, its gauge invariance is ensured.

The QCA operators that will be defined in the next sections are linear combination or exponentials of the hopping, electric and magnetic terms. Therefore, their gauge invariance is ensured by the individual gauge invariance of those terms. 

\subsubsection{Gauge invariant states}

In a gauge theory, gauge transformations should not be observable. Let $\mathcal{O}$ be an observable, $\rho$ a density matrix representing a state and $g_\varphi$ a gauge transformation. That gauge transformations should not be observable amounts to asking that $\Tr(\mathcal{O} g_\varphi \rho g_\varphi^\dagger) = \Tr(\mathcal{O} \rho)$. There are two common ways to enforce this: namely to restrict the set of observables to the gauge invariant ones $[\mathcal{O},g_\varphi]=0$, so that $\Tr(\mathcal{O} g_\varphi \rho g_\varphi^\dagger)=\Tr(g_\varphi\mathcal{O}   \rho g_\varphi^\dagger) = \Tr(g_\varphi^\dagger g_\varphi \mathcal{O}  \rho )= \Tr(\mathcal{O} \rho)$ or to restrict the set of states to the gauge invariant ones, so that $\Tr(\mathcal{O} g_\varphi \rho g_\varphi^\dagger)=\Tr(\mathcal{O} \rho g_\varphi g_\varphi^\dagger) = \Tr(\mathcal{O} \rho)$. We opt for the second, demanding that for every gauge transformation,
\begin{equation}\label{eq:gaugeinvstate}
    g_\varphi \rho  = \rho g_\varphi.
\end{equation} 
Let us draw the consequences of this demand that the states be gauge-invariant. Any density matrix $\rho$ is a convex linear distribution over pure states. In the case of a pure state $\ket{\psi}\bra{\psi}$, this commutation relation amounts to forbidding superposition across any two eigenspaces of a gauge transformation, so that $g_\varphi \ket{\psi}\bra{\psi}= e^{i\lambda}\ket{\psi}\bra{\psi}=\ket{\psi}\bra{\psi}e^{i\lambda}=\ket{\psi}\bra{\psi}g_\varphi$ with $e^{i\lambda}$ the eigenvalue of the eigenspace of $g_\varphi$ to which $\ket{\psi}$ pertains. In other words, $\ket{\psi}=\sum_{c\in S}\alpha_c\ket{c}$ is a superposition of particular basis states $\ket{c}$, i.e. taken in some subset $S$ such that for all $\varphi$ there exists $\lambda$, such that for all $c\in S$, $g_\varphi\ket{c}=e^{i\lambda}\ket{c}$. Let $f(x)$ denote the sum of the occupation numbers, for both the fermions and the gauge fields, at each site $x$ of $c$. Because the $\lambda=\sum_x\varphi(x)f(x)$ and $\varphi(x)$ is arbitrary, the gauge-invariance therefore imposes that $f(x)$ be the same for $c\in S$. Thus, gauge-invariance demands that there exists some fixed, classical occupation number function $f(x)$, and that pure states be considered physical if and only if, at each position $x$, the occupation number operator yields $f(x)$:
\begin{equation}
    \left( \sum_{\eta \in \{\pm \mu, \pm \nu, \pm \kappa\}} E_{x:\eta} + \sum_{j\in 0 \ldots (d-1)} \ket{1}^{x,j}\bra{1} \right) \ket{\psi_\text{phys}} = f(x) \ket{\psi_\text{phys}}.
\end{equation}
The physical pure states then form a subspace, call it the $f(x)$-subspace. The choice of a particular $f(x)$ can be interpreted physically as the choice of a classical fixed, external electromagnetic field. 

Having fixed $f(x)$ and thus $S$, one may wonder about the operators which allow us to prepare some basic state $c'\in S$ given initial basic state $c\in S$, e.g. by creating a fermion. As seen previously, the fermionic annihilators $a_{x,j}$ is not gauge invariant; it does not preserve occupation numbers and will take us out of $S$. Following ideas from \cite{Melnikov2000LatticeSM}, each fermion creation operator $a^\dagger_{x,j}$ could be turned into a gauge invariant state preparation by accompanying it with a gauge field lowering operator $V_{x:\eta}$, but this changes the occupation number at position $x+\eta$, which in turn has to be compensated by a $V_{x+\eta:\zeta}$, and so on until a boundary (possibly infinitely far) is reached. The following defined a gauge invariant state preparation:
\begin{equation}   
    \overline{a}^\dagger_{x,j,p} = a^\dagger_{x,j} \prod_{y:\eta \in p} V_{y:\eta}
\end{equation}
where $p$ is a gauge field path from position $x$ to the space boundary (possibly infinite). If the lattice is finite, the last operator is understood to be a half-link transformation $s$, instead of $V$, since there would be no end to the last link. (Notice that $a^\dagger_{x,j}s_{x:\eta}$ is also gauge invariant operator, but disallowed as it breaks the restriction that $E_{x:\eta}=-E_{x+\eta:-\eta}$, except at the boundary.)

The state preparation $\overline{a}^\dagger_{x,j,p}$ follows the prescribed anti-commutation relations. Indeed, recalling Eq. \eqref{eq:bosonfermioncommutation}, we have that $a_{x,j}$ commutes with any $V_{y:\eta}$, hence $\overline{a}^\dagger_{x,j,p}$ also does. Moreover, within two operators $\overline{a}_{x,j,p}$ and $\overline{a}_{y,k,q}$, the fermionic parts $a_{x,j}$ and $a_{y,k}$ anti-commute, while every other pair of operators commutes, enforcing the fermionic anti-commutation relation \eqref{eq:fermioncommutation}. 


Another gauge invariant state preparation is the creation of a gauge field loop, i.e. $\prod_{y:\eta \in p} V^\dagger_{y:\eta}$ where $p$ is a cyclic gauge field path, or one that begins and ends at a boundary.

\paragraph*{Pair creation.} 
\PA{Our construction is completely general and offers the possibility of working in specific subspaces to mimic physically-relevant phenomena, such the pair production, i.e., the spontaneous creation of electrons and positrons from the QED vacuum \cite{banuls2020simulating, Magnifico2020realtimedynamics}. For this purpose, let us consider the sector in which for all site $x$, $f(x)=d/2$. In this subspace, a natural choice for representing the canonical vacuum is the state in which all the antiparticle  degrees of freedom are populated, whereas no particle and gauge fields are present (this state can be viewed as the reference ``Dirac sea''). For instance in $2+1$ dimensions, when $d=2$, the canonical vacuum would have $\ket{10}^{(x,1)(x,0)}$ at each site $x$, with all gauge fields set to $\ket{0}^{x:\eta}$. Acting on such a state, a term like $a_{x,1}a^\dagger_{x,0}$ yields $\ket{01}^{(x,1)(x,0)}$. This can be understood as `creating particle, namely an electron, at $(x,0)$' at the cost of `creating an antiparticle hole, namely a positron, at $(x,0)$'. This evolution describes the process of electron-positron pair creation from the reference vacuum. More precisely it models the very moment when, at site $x$, the two particles are created : it is only once they move away from each other that the gauge fields will get updated, so as to account for a gauge field line between them two. The reverse evolution describes electron-positron pair annihilation.}

\section{\texorpdfstring{$2+1$}{2+1} Quantum Cellular Automaton}
\label{sec:evolution2d}

Using the previously defined local, gauge invariant operators, it is possible to define a QCA that accounts for QED in $2+1$ dimensions. We proceed in three steps, following the same principles as that leading to the QED Lagrangian. First, a gauge invariant free dynamics for the fermions is defined through a generalization of a Dirac quantum walk (QW) to multiple walkers. Second, the electric contribution is defined as one of the simplest gauge invariant operator acting on the gauge field based on the electric operator, following ideas from \cite{Arrighi2020AQC}. It is found to match the Trotterization of the electric part of the Kogut-Susskind Hamiltonian  \cite{Kogut1975HamiltonianFO}. Third, the magnetic part is added as one the simplest gauge invariant operator acting on the gauge field with the lowering and raising operators. In fact, two constructions are provided for this magnetic term, both agreeing in the limit and matching a Trotterization of the magnetic Kogut-Susskind Hamiltonian.

\subsection{Fermionic dynamics}\label{sec:fermionicdyn}

Let us first recall the $2+1$ dynamics of the fermionic field, without any electromagnetic contribution, without any gauge field even, and restricting to the one particle sector. This is the well-known Dirac quantum walk (QW) \cite{di2012discrete,Arrighi2014TheDE}.  

At each lattice site $x$ lies a group of $2$ qubits, each stating whether a fermion in mode $j\in 0\ldots 1$ is present at the site (remember that in $2+1$ dimensions the Dirac Eq. is a PDE on a wave function having two complex amplitudes). Again, this encoding captures the Pauli exclusion principle as there cannot be two fermions in the same mode at the same site. Moreover, as we focus on the one particle sector first, we temporarily look at the case where only one of the qubits can be in the state $\ket{1}$.

One time-step is divided in three sub steps: a vertical translation, a horizontal translation and a mass term. Moreover, each translation is decomposed into two swaps. Each of these terms thus acts on a pair of qubits: on a single site for the mass and the first swap ($S$), and across two neighbouring sites for the second swaps ($T$). In the one particle sector, the possible input states are $\ket{00}$, $\ket{01}$ and $\ket{10}$. Number conservation forces $\ket{00}$ to be mapped to itself. Without loss of generality we can assume that this absence of particle triggers no phase. \PA{Hence, our on-site (or across-two-sites), two qubit operators for the Dirac QW are of the form $W=1\oplus M$ where $1$ leaves $\ket{00}$ unchanged and $M$ is the unitary acting on the subspace spanned by $\ket{01}$ and $\ket{10}$. The Dirac QW is a global operator expressed in terms of a large product of such on-sites operators (which individually are independent of $x$):}
\begin{align}
    \text{QW} &=  \left[\bigotimes_x C_\epsilon \right] \left[\bigotimes_{(x,1),(x+\nu,0)} T_\nu  ~~\bigotimes_x S  \right] \left[\left(\bigotimes_x H_\mu \right) \left(\bigotimes_{(x,1),(x+\mu,0)} T_\mu  ~~\bigotimes_x S \right) \left(\bigotimes_x H_\mu^\dagger \right) \right]
\end{align}
where $C_\epsilon = 1\oplus e^{-im\epsilon Y}$ is the mass term, $H_\mu = 1\oplus H$ with $H$ the Hadamard operator (such that $HZ H =X$), $S$ swaps qubits $(x,0)$ and $(x,1)$, and $T_\eta$ swaps qubits $(x,1)$ and $(x+\eta,0)$. It follows that $\bigotimes_{(x,1),(x+\eta,0)} T_\eta  \bigotimes_x S = 1\oplus e^{\epsilon Z \partial_\eta}$ is the displacement operators in the $\eta$ direction by a factor $\epsilon$, moving the first qubit in the positive direction and the second in the negative direction. The $X, Y, Z$ are Pauli matrices and $\partial_\eta$ the partial derivatives. Convergence towards the Dirac Eq. is rigorously proven in \cite{di2012discrete,Arrighi2014TheDE}. \PA{Again this was without any electromagnetic contribution, without any gauge field even, and restricting to the one particle sector.}

Moving on to multiple walkers, the Dirac QW becomes the Dirac QCA. The dynamics needs to be extended to take into account the case where multiple qubits are in state $\ket{1}$. Since the QW on-sites operator acted on at most two qubits, and due to number conservation, only the input state $\ket{11}$ requires our attention, and it can only be sent to itself, up to a phase. To find out exactly which phase has to be applied, let us move to the Heisenberg picture. 

The Heisenberg picture tells about the future impact of our past actions. For instance, say that the overall evolution from time $t$ to time $t+1$ is governed by a unitary operator $\mathbf{W}$, e.g. mapping $\ket{\psi}$ to $\ket{\psi'}$. Then, the past action of creating a fermion at $(x,0)$ at time $t$, as implemented by $a^\dagger_{x,0}$, e.g. mapping  $\ket{\psi}$ to $a^\dagger_{x,0}\ket{\psi}$, will have future impact $\mathbf{W} a^\dagger_{x,0} \mathbf{W}^\dagger$ at time $t+1$, i.e. $\ket{\psi'}$ to $\mathbf{W} a^\dagger_{x,0} \mathbf{W}^\dagger\ket{\psi'}$. 

More specifically consider $\mathbf{S}$ the multi-particle sector extension of $S$ which is such that $\mathbf{S}_x a^\dagger_{x,0} \mathbf{S}_x^\dagger = a^\dagger_{x,1}$ and $\mathbf{S}_x a^\dagger_{x,1} \mathbf{S}_x^\dagger = a^\dagger_{x,0}$, with $\mathbf{S}_x$ acting as $\mathbf{S}$ on site $x$ and as the identity elsewhere. Such an $\mathbf{S}$ is called `fermionic swap'. Notice there exists $\mathbf{S}'$ which coincide with $\mathbf{S}$ (and thus with $S$) in the one-particle sector, but that do not obey these two equations.   We discard them on the basis that $\mathbf{S}$ is the `rightful non-interacting extension' of $S$, as the future impact of $a^\dagger_{x,0}$ ought to be $a^\dagger_{x,1}$ regardless of there being other particles or not. 

It follows that $\mathbf{S}_x a^\dagger_{x,0}a^\dagger_{x,1} \mathbf{S}_x^\dagger = \mathbf{S}_x a^\dagger_{x,0}\mathbf{S}_x\mathbf{S}_x^\dagger a^\dagger_{x,1} \mathbf{S}_x^\dagger=a^\dagger_{x,1}a^\dagger_{x,0}=-a^\dagger_{x,0}a^\dagger_{x,1}$. In particular, if we had $\ket{\psi}^{(x,1)(x,0)}=\ket{00}^{(x,1)(x,0)}$ at time $t$ evolving into $\ket{\psi'}^{(x,1)(x,0)}=\mathbf{S}\ket{\psi}^{(x,1)(x,0)}=\ket{00}^{(x,1)(x,0)}$ at time $t+1$, we realize that the past action of creating two fermions $a^\dagger_{x,0}a^\dagger_{x,1}$, e.g. mapping $\ket{00}^{(x,1)(x,0)}$ to $\ket{11}^{(x,1)(x,0)}$, will have future impact $-a^\dagger_{x,0}a^\dagger_{x,1}$, i.e. mapping $\ket{00}^{(x,1)(x,0)}$ to $-\ket{11}^{(x,1)(x,0)}$. Thus, $\mathbf{S}\ket{11}^{(x,1)(x,0)}=-\ket{11}^{(x,1)(x,0)}$ is the rightful qubit implementation of the fermionic swap, meeting the specifications imposed by the (anti\nobreakdash-)commutation relations hypothesis. In order to build the Dirac QCA from the Dirac QW, we must proceed in the same manner for the different on-sites operator making up the Dirac QW, as done in Appendix \ref{appendix:qwtoqca}.

We see that this phase got determined as a consequence of the (anti-)commutation relations hypothesis discussed in Sec. \ref{sec:jordanwignerext}, as well as the way we chose to implement the annihilation operators, under which Jordan-Wigner transform etc., as defined in that same section. This requires the presence of gauge field, which we had temporarily ignored for describing the Dirac QW, but we now restore for describing the Dirac QCA. We use bold fonts to denote the Dirac QCA counterparts of the Dirac QW operators. 

\paragraph{The mass sub-step} acts at each site $x$ with on-site operator $C_\epsilon$ over $\ket{00}^{(x,1)(x,0)}$, $\ket{01}^{(x,1)(x,0)}$ and $\ket{10}^{(x,1)(x,0)}$. It applies a phase to the state $\ket{11}^x$, which is just $c^2+ s^2=1$ (using Appendix \ref{appendix:qwtoqca}). Therefore, the corresponding QCA on-site operator $\mathbf{C}_\epsilon$ is: 
\begin{align}\label{eq:2dmass}
    \mathbf{C}_\epsilon =& \begin{pmatrix}
        1 & 0 & 0 & 0 \\
        0 & c & -s & 0 \\
        0 & s & c & 0 \\
        0 & 0 & 0 & 1 
    \end{pmatrix}\\
    =& C_\epsilon \oplus 1
\end{align}
\PA{where the matrix is expressed as acting on some $\ket{\psi}^{(x,1)(x,0)}$, i.e. with respect to the canonical basis $\{\ket{00}^{(x,1)(x,0)},\ket{01}^{(x,1)(x,0)},\ket{10}^{(x,1)(x,0)},\ket{11}^{(x,1)(x,0)} \}$, in this order.} 
Let $\mathbf{C}_{x,\epsilon}$ be the local operator which acts as $\mathbf{C}_\epsilon$ on site $x$, and as the identity elsewhere. We have
\begin{align}
    \mathbf{C}_{x,\epsilon}=& \left(a_{x,1} a^\dagger_{x,1} \right)\left( a_{x,0} a^\dagger_{x,0}\right) + \left(a^\dagger_{x,1} a_{x,1} \right)\left( a^\dagger_{x,0}a_{x,0}  \right) \\
     & + c \left[\left(a_{x,1} a^\dagger_{x,1} \right)\left(a^\dagger_{x,0} a_{x,0}\right) + \left(a^\dagger_{x,1} a_{x,1}\right)\left( a_{x,0} a^\dagger_{x,0}\right) \right]+ s (a^\dagger_{x,1} a_{x,0} - a^\dagger_{x,0} a_{x,1} )
\end{align}
where $c=\cos(\epsilon m)$ and $s=\sin(\epsilon m)$, and the last line is $\mathbf{C_\epsilon}$ expressed in terms of the local evolution operator from Eq. \eqref{eq:massterm}.

\paragraph{The transport sub-step} takes a right-moving (resp. left-moving) qubit at position $x$ (resp. $x+\eta$) and maps it to the right-moving (resp. left-moving) qubit at position $x+\eta$ (resp. $x$). 
It does so by first swapping the qubits on each site using the on-site operator $\mathbf{S}$, and then moving them through an across-two-sites operator $\mathbf{T_{\eta}}$, whilst changing the gauge field accordingly. \PA{Here is $\mathbf{S}$ as acting on some $\ket{\psi}^{(x,1)(x,0)}$},
\begin{align}\label{eq:swap}
    \mathbf{S}=&\begin{pmatrix}
        1 & 0 & 0 & 0 \\
        0 & 0 & 1 & 0 \\
        0 & 1 & 0 & 0 \\
        0 & 0 & 0 & -1 
    \end{pmatrix}
\end{align}
Thus,
\begin{align*}
    \mathbf{S}_x=& \left(a_{x,1} a^\dagger_{x,1} \right) \left( a_{x,0} a^\dagger_{x,0} \right) - \left(a^\dagger_{x,1} a_{x,1} \right) \left( a^\dagger_{x,0}a_{x,0} \right) + a^\dagger_{x,1} a_{x,0} + a^\dagger_{x,0} a_{x,1} 
\end{align*}
\PA{Here is $\mathbf{T}_\eta$ as acting on some $\ket{\psi}^{(x+\eta,0)(x,1)}$},
\begin{align}
    \label{eq:transport}
    \mathbf{T_{\eta}}=&
    \begin{pmatrix}
        1 & 0 & 0 & 0 \\
        0 & 0 & K_\eta & 0 \\
        0 & K^\dagger_\eta & 0 & 0 \\
        0 & 0 & 0 & -1 
    \end{pmatrix}
\end{align}
where $K_{\eta}$ is
\begin{align}
    K_{\eta}=
    \left(\bigotimes_{y\in \llbracket  (x,j), x:\eta \llbracket } Z_y \right) 
    U_{x:\eta}
    \left(\bigotimes_{y\in \llbracket  (x+\eta,k), x+\eta:-\eta \llbracket } Z_y \right) 
\end{align}
with $j=1$, $k=0$, and $U_{x:\eta}$ the action of the gauge field lowering operator on $(x:\eta)(x+\eta:-\eta)$. Let $\mathbf{T}_{x,\eta}$ be the local operator which acts as $\mathbf{T}_\eta$ across sites $x$ and $x+\eta$ and as the identity elsewhere. We have
\begin{align}
    \mathbf{T}_{x,\eta}=& \left( a_{x,1} a^\dagger_{x,1} \right) \left(  a_{x+\eta,0} a^\dagger_{x+\eta,0} \right) - \left(a^\dagger_{x,1} a_{x,1} \right) \left( a^\dagger_{x+\eta,0}a_{x+\eta,0}\right)  \\
    & + a^\dagger_{x,1}V_{x:\eta} a_{x+\eta,0} + a^\dagger_{x+\eta,0}V^\dagger_{x:\eta} a_{x,1}
\end{align}
 The equations using fermionic annihilators and creators are based on the local evolution operators from Eqs. \eqref{eq:massterm} and \eqref{eq:hoppingterm}. The product of $Z_y$ comes from the hopping term defined in Eq. \eqref{eq:localhoppingterm}, with $x:\eta$ the link along which the swap takes place. The minus one, when the input qubits are in state $\ket{11}^{(x+\eta,0)(x,1)}$, is the exchange phase for crossing fermions. Again full justification is given in Appendix \ref{appendix:qwtoqca}.

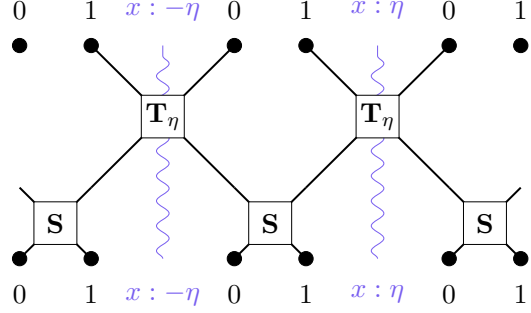
\begin{figure*}[ht!]
    \centering
    \resizebox{0.5\textwidth}{!}{\newcommand{\colora}{\cba}
\newcommand{\colorb}{\cbb}
\newcommand{\colorc}{\cbc}

\begin{tikzpicture}

\newcommand{\state}[2]{\filldraw[color=black, fill=black] (#1,#2) circle (0.1);}
\newcommand{\states}[2]{
    \begin{scope}[shift={(#1,#2)}] 
        \state{0}{0};
        \state{1}{0};
    \end{scope}
}
\newcommand{\naming}[4]{
    \begin{scope}[shift={(#1,#2)}] 
        \node at (0,0) {#3};
        \node at (1,0) {#4};
    \end{scope}
}

\newcommand{\gate}[4]{
\begin{scope}[shift={(#1,#2)}] 
    \draw[thick] (1-#3,1-#3) -- (#3,#3) (#3,1-#3) -- (1-#3,#3);
    \filldraw[color=black, fill=white] (0.2,0.2) rectangle (0.8,0.8);
    \node at (.5,.5) {#4};
\end{scope}}
\newcommand{\swap}[2]{\gate{#1}{#2}{1}{$\mathbf{S}$}}
\newcommand{\transport}[2]{\gate{#1}{#2}{1.5}{$\mathbf{T_{\eta}}$}}

\states{-3}{0}
\naming{-3}{-.5}{0}{1}%
\states{0}{0}
\naming{0}{-.5}{0}{1}%
\states{3}{0};
\naming{3}{-.5}{0}{1}%

\node[color=\colorb] at (-1,-.5) {$x:-\eta$};%
\node[color=\colorb] at (2,-.5) {$x:\eta$};%
\path[draw=\colorb,snake it] (-1,0) -- (-1,3);
\path[draw=\colorb,snake it] (2,0) -- (2,3);

\transport{-1.5}{1.5}
\transport{1.5}{1.5}
\swap{-3}{0}
\swap{0}{0}
\swap{3}{0}

\begin{scope}[shift={(0,3)}] 
    \states{-3}{0};
    \naming{-3}{.5}{0}{1}%
    \states{0}{0}
    \naming{0}{.5}{0}{1}%
    \states{3}{0};
    \naming{3}{.5}{0}{1}%
    \node[color=\colorb] at (-1,.5) {$x:-\eta$};%
    \node[color=\colorb] at (2,.5) {$x:\eta$};%
\end{scope}

\end{tikzpicture}}
    \caption{Transport with qubit $(x,0)$ moving right (and $(x,1)$ moving left) and $T$ updating the gauge fields accordingly (as represented as a single wiggly line for conciseness).}
    \label{fig:transport}
\end{figure*}

The transport sub-step of the Dirac QCA is illustrated in Fig. \ref{fig:transport}.

\paragraph{The basis change}$\mathbf{H}$ is similar to the mass term in that it is an on-site operator that can be written as a direct sum for the case with $0$, $1$ or $2$ particles as follows (\PA{as acting on some $\ket{\psi}^{(x,1)(x,0)}$}):
\begin{align}\label{eq:basis2}
    \mathbf{H} &= H_\mu \oplus -1 =1 \oplus  H \oplus -1 \\
        &= \begin{pmatrix}
            1 & 0 & 0 & 0 \\
            0 & \frac{1}{\sqrt2} & \frac{1}{\sqrt2} & 0 \\
            0 & \frac{1}{\sqrt2} & -\frac{1}{\sqrt2} & 0 \\
            0 & 0 & 0 & -1 \\
        \end{pmatrix}
\end{align}
Let  $\mathbf{H}_x$ be the local operator acting as $\mathbf{H}$ on site $x$ and as the identity elsewhere. We have
\begin{align}            
        \mathbf{H}_x=& \left(a_{x,1} a^\dagger_{x,1} \right)\left( a_{x,0} a^\dagger_{x,0}\right) - \left(a^\dagger_{x,1} a_1 \right)\left( a^\dagger_{x,0}a_{x,0}  \right) \\
         & + \frac{1}{\sqrt2} \left[\left(a_{x,1} a^\dagger_{x,1} \right)\left(a^\dagger_{x,0} a_{x,0}\right) + \left(a^\dagger_{x,1} a_{x,1}\right)\left( a_{x,0} a^\dagger_{x,0}\right) + (a^\dagger_{x,1} a_{x,0} - a^\dagger_{x,0} a_{x,1} )\right].
\end{align}
This last equation corresponds to the basis change expressed using the local evolution operator \eqref{eq:massterm};

The minus one is justified in Appendix \ref{appendix:qwtoqca}. Since $H^\dagger = H$, we have $\mathbf{H^\dagger} = \mathbf{H}$.

The complete Dirac QCA is:

\PA{
\begin{align}
    \mathbf{D_F} &=  \left[\bigotimes_x \mathbf{C}_\epsilon \right] \left[\bigotimes_{(x,1),(x+\nu,0)} \mathbf{T}_\nu  ~~\bigotimes_x \mathbf{S}  \right] \left[\left(\bigotimes_x \mathbf{H}_\mu \right) \left(\bigotimes_{(x,1),(x+\mu,0)} \mathbf{T}_\mu  ~~\bigotimes_x \mathbf{S} \right) \left(\bigotimes_x \mathbf{H}_\mu^\dagger \right) \right]\label{eq:qcafermion}
\end{align}
}

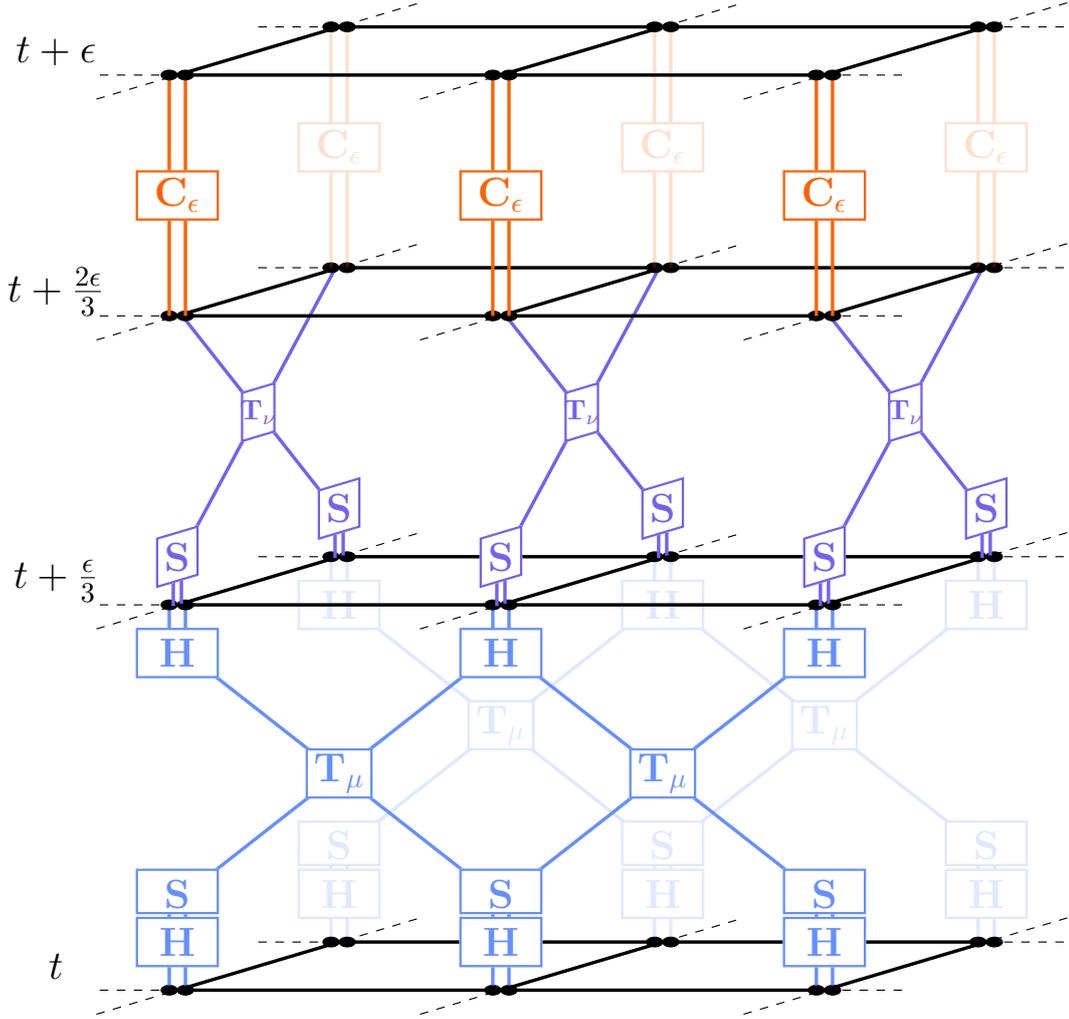
\begin{figure*}[ht!]
    \centering
    \Large\resizebox{\textwidth}{!}{\newcommand{\colora}{\cba}
\newcommand{\colorb}{\cbb}
\newcommand{\colorc}{\cbd}
\newcommand{\colord}{\cbc}
\begin{tikzpicture}

\begin{scope}[yscale=0.6] 

\newcommand{\state}[2]{
\filldraw[color=black, fill=black] (#1-0.1,#2) circle (0.1);
\filldraw[color=black, fill=black] (#1+0.1,#2) circle (0.1);
}
\newcommand{\grid}{
\state{0}{0}
\state{2}{1}
\state{4}{0}
\state{6}{1}
\state{8}{0}
\state{10}{1}
\draw[color=black, very thick] (0,0) -- (2,1) -- (6,1) -- (4,0) -- (0,0);
\draw[color=black, very thick] (6,1) -- (10,1) -- (8,0) -- (4,0);
\draw[color=black, dashed] (-1,-0.5) -- (0,0) -- (-1,0)
                           (1,1) -- (2,1) -- (3,1.5)
                           (3,-.5) -- (4,0)
                           (6,1) -- (7,1.5)
                           (7,-.5) -- (8,0) -- (9,0)
                           (11,1.5) -- (10,1) -- (11,1);
}

\begin{scope}[shift={(0,-1)}] 
    \grid
    \node at (-1.5,0.5) {$t$};
\end{scope}
\newcommand{\basis}[2]{
    \draw[color=#1, very thick, opacity=#2] (0.1,-.9) -- (.1,.6) (-0.1,-.9) -- (-.1,.6);
    \draw[color=#1, very thick, opacity=#2] (0.1,6.5) -- (.1,7) (-0.1,6.5) -- (-.1,7);
    \filldraw[color=white, fill=white, thick] (-0.5,-.5) rectangle (0.5,.5);
    \filldraw[color=#1, fill=white, thick,  opacity=#2] (-.5,-.5) rectangle (.5,.5);
    \node[color=#1, opacity=#2] at (0,0) {$\mathbf{H}$};
    \filldraw[color=white, fill=white, thick] (-0.5,0.6) rectangle (0.5,1.5);
    \filldraw[color=#1, fill=white, thick,  opacity=#2] (-.5,0.6) rectangle (.5,1.5);
    \node[color=#1, opacity=#2] at (0,1) {$\mathbf{S}$};
    \filldraw[color=white, fill=white, thick] (-0.5,5.5) rectangle (0.5,6.5);
    \filldraw[color=#1, fill=white, thick,  opacity=#2] (-.5,5.5) rectangle (.5,6.5);
    \node[color=#1, opacity=#2] at (0,6) {$\mathbf{H}$};
}
\newcommand{\xevo}[4]{
\begin{scope}[shift={(#1,#2)}] 
    \draw[color=#3, very thick, opacity=#4] (0.1,1) -- (3.9,6) (3.9,1) -- (0.1,6);
    \filldraw[color=white, fill=white, thick] (1.6,3) rectangle (2.4,4);
    \filldraw[color=#3, fill=white, thick,  opacity=#4] (1.6,3) rectangle (2.4,4);
    \node[color=#3, opacity=#4] at (2,3.5) {$\mathbf{T}_{\mu}$};
    \basis{#3}{#4}
\end{scope}
}
\xevo{0}{0}{\colora}{1}
\xevo{4}{0}{\colora}{1}
\xevo{2}{1}{\colora}{0.2}
\xevo{6}{1}{\colora}{0.2}
\begin{scope}[shift={(8,0)}]\basis{\colora}{1}\end{scope}
\begin{scope}[shift={(10,1)}]\basis{\colora}{.2}\end{scope}

\begin{scope}[shift={(0,7)}] 
    \grid
    \node at (-1.5,0.5) {$t+\frac{\epsilon}{3}$};
\end{scope}

\newcommand{\yswap}[2]{
    \draw[color=#1, very thick, opacity=#2] (.1,0) -- (.1,.5) (-0.1,0) -- (-.1,.5);
    \filldraw[color=white, fill=white, thick] (-0.5,.5) rectangle (0.5,1.5);
    \filldraw[color=#1, fill=white, thick,  opacity=#2] (-.5,.5) rectangle (.5,1.5);
    \node[color=#1, opacity=#2] at (0,1) {$\mathbf{S}$};
}
\newcommand{\yevo}[4]{
\begin{scope}[shift={(#1,#2)}, xscale=0.5,yslant=0.25] 
    \draw[color=#3, very thick, opacity=#4] (0.1,1) -- (3.9,6) (3.9,1) -- (0.1,6);
    \filldraw[color=white, fill=white, thick] (1.6,3) rectangle (2.4,4);
    \filldraw[color=#3, fill=white, thick,  opacity=#4] (1.6,3) rectangle (2.4,4);
    \node[color=#3, opacity=#4] at (2,3.5) {\small$\mathbf{T}_{\nu}$};
\end{scope}
}
\yevo{0}{7}{\colorb}{1}
\yevo{4}{7}{\colorb}{1}
\yevo{8}{7}{\colorb}{1}
\begin{scope}[shift={(0,7)},xscale=0.5,yslant=0.25]
    \yswap{\colorb}{1}
\end{scope}
\begin{scope}[shift={(4,7)},xscale=0.5,yslant=0.25]
    \yswap{\colorb}{1}
\end{scope}
\begin{scope}[shift={(8,7)},xscale=0.5,yslant=0.25]
    \yswap{\colorb}{1}
\end{scope}
\begin{scope}[shift={(2,8)},xscale=0.5,yslant=0.25]
    \yswap{\colorb}{1}
\end{scope}
\begin{scope}[shift={(6,8)},xscale=0.5,yslant=0.25]
    \yswap{\colorb}{1}
\end{scope}
\begin{scope}[shift={(10,8)},xscale=0.5,yslant=0.25]
    \yswap{\colorb}{1}
\end{scope}

\begin{scope}[shift={(0,13)}] 
    \grid
    \node at (-1.5,0.5) {$t+\frac{2\epsilon}{3}$};
\end{scope}

\newcommand{\mass}[4]{
\begin{scope}[shift={(#1,#2)}] 
    \draw[color=#3, very thick, opacity=#4] (0.1,0) -- (0.1,5);
    \draw[color=#3, very thick, opacity=#4] (-.1,0) -- (-0.1,5);
    \filldraw[color=white, fill=white, thick] (-.5,2) rectangle (.5,3);
    \filldraw[color=#3, fill=white, thick,  opacity=#4] (-.5,2) rectangle (.5,3);;
    \node[color=#3, opacity=#4] at (0,2.5) {$\mathbf{C}_\epsilon$};
\end{scope}
}
\mass{0}{13}{\colorc}{1}
\mass{4}{13}{\colorc}{1}
\mass{8}{13}{\colorc}{1}
\mass{2}{14}{\colorc}{.2}
\mass{6}{14}{\colorc}{.2}
\mass{10}{14}{\colorc}{.2}

\begin{scope}[shift={(0,18)}] 
    \grid
    \node at (-1.5,0.5) {$t+\epsilon$};
\end{scope}

\end{scope}
\end{tikzpicture}}
    \caption{The 3 steps of the free evolution in the multi-particle sector. The gauge field is omitted for clarity.}\label{fig:fermionicevolution}
\end{figure*}

It is represented in Fig. \ref{fig:fermionicevolution}.

\paragraph{Gauge invariance.} Each of the operators constitutive of the Dirac QCA has been expressed as a linear combination of the local evolution operators given in Subsec. \ref{sec:locality}, which were proven to be gauge invariant in Subsec. \ref{sec:gaugeinv}. Therefore, the Dirac QCA is gauge invariant.

\subsection{Electric contribution}

Let us now define the electric contribution. To do so, we follow the same idea as in the Lagrangian formalism, which is to take some of the simplest gauge invariant electric term. We then check that it matches the Trotterization of the Kogut-Susskind Hamiltonian. The construction proposed here is highly inspired by \cite{Arrighi2020AQC}.

A simple electric contribution ought to act on the gauge field according to a functional of the local electric counting operator $E_{x:\eta}$ for $\eta\in\{\mu,\nu\}$, as defined in \eqref{eq:electric_op}. We could also demand invariance under $E_{x+\eta:-\eta}$ instead. However, the electric operator itself does not match this requirement because $E_{x:\eta}= -E_{x+\eta:-\eta}$. The squared electric term $E_{x:\eta}^2$, does. The squared electric term itself is not unitary, but its exponential $e^{iE_{x:\eta}^2}$, is. 

Notice also that the spacetime discretization should impact the amplitude of the phase: when any of the space or the time discretization parameter goes to zero, the exponential should tend towards the identity. Taking these considerations into account, one obtains:
\begin{equation} \label{eq:qcaelectric}
    \mathbf{D_E} = \bigotimes_{x,\eta\in\{\mu,\nu\}} e^{\frac{i}{2}\epsilon^2 g_E^2 E_{x:\eta}^2}
\end{equation} 
where the $\epsilon^2$ comes from the simultaneous discretization in space and time with $\Delta_x = \Delta_t = \epsilon$, and $g_E$ is the coupling constant as determined experimentally. 

Without the electric contribution, the gauge field just keeps track of fermions passing by, but has no influence upon them. With the electric contribution, the gauge-field dependent phase, mediates the interaction.

Let us compare the electric contribution with the electric part of the Kogut-Susskind Hamiltonian \cite{Kogut1975HamiltonianFO} for QED:
\begin{equation}
    \mathcal{H}_E = \frac{g_E^2}{2}\Delta_x\sum_{x,\eta\in\{\mu,\nu\}}{(E^2_{x:\mu}+E^2_{x:\nu})}.
\end{equation}
Integration over a $\Delta_t$ period of time---i.e. computing $e^{i\Delta_t \mathcal{H}_E}$---and Trotterizing to separate the spatial sum into a product of exponentials, yield $\mathbf{D_E}$.

\begin{remark}[Truncation of the electric field]\label{rem:truncationE}
    If $\frac{1}{2}\epsilon^2 g_E^2 =\frac{2\pi}{k}$ exactly, with $k$ an integer, then the phase of the electric contribution will wrap up around $2\pi$ as soon as $E^2$ reaches $k$. This happens when $\epsilon = \sqrt{\frac{4\pi}{kg_E^2}}$. If we restrict ourselves to such values of $\epsilon$, i.e. decreasing it by augmenting $k$, then, as far as the electric contribution is concerned, the gauge field can equally well be represented with a $k$-dimensional Hilbert space $\mathcal{H}_k$. Indeed, when $E$ outputs $k+l$, its square gives $k^2+2kl + l^2$, and the phase $\frac{1}{2}\epsilon^2 g_E^2 E^2$ simplifies into $\frac{1}{2}\epsilon^2 g_E^2 l^2$ because $\epsilon^2 g_E^2 k(k+2l)/2$ is proportional to $2\pi$. Notice that this $k$ still goes to infinity when taking $\epsilon$ to zero, augmenting proportionally to $1/\epsilon^2$. This idea of restricting the gauge field to finite-dimensions labelling roots of unity was suggested in \cite{Arrighi2020AQC}. Truncations in lattice gauge theory simulations were evaluated in \cite{ercolessi2018phase}.
\end{remark}

\paragraph*{Gauge invariance.} The electric contribution is gauge invariant as an exponential of a gauge invariant operator.

\subsection{Magnetic contribution}
\label{sec:magcontrib}

In order to define the magnetic contribution, we follow the same path as for the electric contribution, that is to say define some of the simplest gauge field-only term that is local, unitary and gauge invariant. 

\paragraph*{Formulation of the magnetic contribution.} \PA{The magnetic contribution acts by lowering or raising the gauge field.}

As discussed in Sec. \ref{sec:jordanwignerext} operators that are defined solely using the lowering and raising operators need to form a loop (spatially) in order to both be local and ensure gauge invariance. Indeed, raising the gauge field at one end of a link, implies lowering it at the other, hence imposing that the magnetic contribution be a loop; the smallest loop is realized by the plaquette local evolution operator $P_{x:\eta,\zeta}$ of Eq. \eqref{eq:gaugelocality}. 

Since the loop is oriented, one may wish to symmetrize. This is done through a sum $P_{x:\eta,\zeta}+P_{x:\eta,\zeta}^\dagger$, but this breaks unitarity. Just like for the electric field, unitarity is restored through exponentiation, and the space and time discretization parameters need be used in the exponential to ensure that it goes to the identity when these go to zero. Taking these considerations into account, one obtains:
\begin{equation} \label{eq:qcamag1}
    \mathbf{D_M} = \bigotimes_{x} e^{\frac{i}{2}\epsilon^2 g_M^2 (P_{x:\mu,\nu} + P_{x:\mu,\nu}^\dagger)}
\end{equation} 
with $\epsilon = \Delta_t = \Delta_x$ and $g_M^2/2$ a constant. Taking the limit of $\mathbf{D_M}$ when $\epsilon$ goes to zero yields:
\begin{equation}\label{eq:qcamag1limit}
    \mathbf{D_M} = Id + i\epsilon^2 \frac{g_M^2}{2} \sum_x\left(P_{x:\mu,\nu} + P_{x:\mu,\nu}^\dagger\right) + O \left(\epsilon^4\right).
\end{equation}

Let us compare the obtained magnetic contribution with magnetic part of the Kogut-Susskind Hamiltonian:
\begin{equation}\label{eq:magnetichamiltonian}
    \mathcal{H}_m = \frac{g_M^2}{2}\Delta_x\sum_{x} (P_{x:\mu,\nu} + P^\dagger_{x:\mu,\nu})
\end{equation}
where $g_M = \frac{1}{\Delta_x g_E}$. Integrating this Hamiltonian over a $\Delta_t$ period of time---i.e. computing $e^{i\Delta_t \mathcal{H}_M}$---and Trotterizing to separate the spatial sum leads to the same local operator $\mathbf{D_M}$.

\paragraph*{Two formulations in terms of gates.} In order to define $\mathbf{D_M}$ in terms of quantum gates, one needs to explicitly compute or closely approximate the exponential. We provide two constructions in order to do so. The first construction is to diagonalize the plaquette term so that taking its exponential simply amounts to exponentiating the eigenvalues, that is to say, we go from an electric basis to a magnetic one \cite{Haase2021ARE}. The quantum Fourier transform is used in the diagonalization. The second construction consist in a more subtle approach were the term $P_{x:\mu,\nu}+P^\dagger_{x:\mu,\nu}$ is written as the sum of two terms $\widetilde{P}_{x:\mu,\nu}$ and $\widetilde{Q}_{x:\mu,\nu}$ whose exponentials are simple to compute. This second construction yields a gate formalism reminiscent to that of a QW.

\paragraph*{Ordering the operators.} Two neighbouring plaquette local evolution operators $P_{x:\mu,\nu}$ and $P_{x+\mu:\mu,\nu}$ both act on the two gauge fields of the link between sites $x+\mu$ and $x+\mu+\nu$. Therefore, the order of the operations could have been relevant. However, the plaquette local evolution operators actually commute. If one insists on not acting simultaneously with two different operators on a same system, then any arbitrary ordering can be chosen, such as applying the plaquette local evolution operators at even positions ($x_1+x_2+x_3~\mod~2=0$) first, and then at odd positions. 

\paragraph*{Redefining the states.} Before diving into the two constructions, let us introduce a new way of representing the gauge field around a plaquette. Forgetting to denote the second gauge field of each link, which is just the opposite of the first, the gauge fields of the four links of the plaquette are
\newcommand{\ncol}{{\color{\cbc} n}}
\newcommand{\npocol}{{\color{\cbc} (n+1)}}
\begin{align*}
&\ket{-a-\ncol,-b-\ncol,c+\ncol,\ncol}^{(x:\mu)(x+\mu: \nu)(x+\nu: \mu)(x: \nu)}
\end{align*}
Let us introduce `plaquette-like' states:
\begin{equation}
    \ket{\underline{abcn}} = \ket{-a-\ncol,-b-\ncol,c+\ncol,\ncol}
\end{equation}
and see how these may be affected by a loop of lowering operators, first using $U$ operators of \eqref{eq:gaugelinklowering_op} instead of the $V$, i.e. 
$$U_{x:\mu} U_{x+\mu: \nu} U_{x+\nu: \mu}^\dagger U_{x: \nu}^\dagger.$$ 
This will just act as a shift on the fourth number 
\begin{equation}
     \ket{\underline{abcn}} \mapsto \ket{\underline{abc,n+1}}  \label{eq:almostplaquettestates}
\end{equation}
and as the identity elsewhere. 

The plaquette local evolution operator of \eqref{eq:vasuz} and \eqref{eq:plaquetteterm}, is a loop of $V$ however, i.e. a loop of $U$ combined with $Z$ operators. These $Z$ operators only add a plus or minus sign to the state $\ket{\underline{abcn}}$. This sign may be influenced by the other gauge fields and qubits present at the four sites where the plaquette term is applied, but not beyond, \PA{cf. the locality stated in \eqref{eq:gaugelocality}. In order to be able to determine this sign, we extend the plaquette-like states to encompass the states of all the qubits and gauge fields of the plaquette that were not already accounted for in $\ket{\underline{abcn}}$. More specifically, we now take as plaquette states $\ket{\underline{abcno}}$ where $\ket{o}$ is a canonical basis state of the systems
\begin{align*}
\{x+\eta:\eta'\}_{\eta\in \{0,\mu,\nu,\mu+\nu\},\ \eta'\in \{\pm\mu,\pm\nu\},\ \eta+\eta'\notin\{0,\mu,\nu,\mu+\nu\}} \cup \ \{(x+\eta,j)\}_{\eta\in \{0,\mu,\nu,\mu+\nu\},j\in\{0,1\}}.
\end{align*}}
We can then let $\varphi_{abcno}\in\{0,1\}$ be such that
\begin{equation}
    P \ket{\underline{abcno}}= (-1)^{\varphi_{abcno}}\ket{\underline{abc,n+1,o}}
\end{equation}
where $P$ is the on-plaquette operator such that $P_{x:\mu,\nu}$ acts as $P$ on the four sites of the $x:\mu,\nu$ plaquette, and as the identity elsewhere.  
Starting from state $\ket{\underline{abc,0,o}}$, one reaches state $\ket{\underline{abc,n,o}}$ by applying the plaquette operator $n$ times. We can then let $\psi_{abcno}\in\{0,1\}$ be such
\begin{equation}
    (-1)^{\psi_{abcno}}\ket{\underline{abcno}}= P^n\ket{\underline{abc,0,o}}.
\end{equation} 
The $\psi_{abcno}$ can be given explicitly in terms of $\varphi_{abcno}$, so that $\psi_{abcno} + \varphi_{abcno} = \psi_{abc,n+1,o}$:
\begin{equation}
    {\psi_{abcno}} = 
    \begin{cases}
        &{\sum_{0\leq k < n} \varphi_{abcko}} \quad \text{when}~n>0\\
        &{\sum_{n\leq k < 0} \varphi_{abcko}} \quad \text{when}~n<0
    \end{cases}
\end{equation}
This leads to the definition of `plaquette states'
\begin{equation}
    \ket{\widetilde{abcno}} = (-1)^{\psi_{abcno}}\ket{\underline{abcno}}
\end{equation}
which verify the same relation as that of Eq. \eqref{eq:almostplaquettestates}, with $V$ operators instead of $U$:
\begin{equation}\label{eq:plaquetteasshift}
    P \ket{\widetilde{abcno}}= \ket{\widetilde{abc,n+1,o}}.
\end{equation}
Again, these plaquette states help understand the plaquette local evolution operator as a shift, which will be useful for both decompositions of the magnetic contribution in terms of quantum gates.

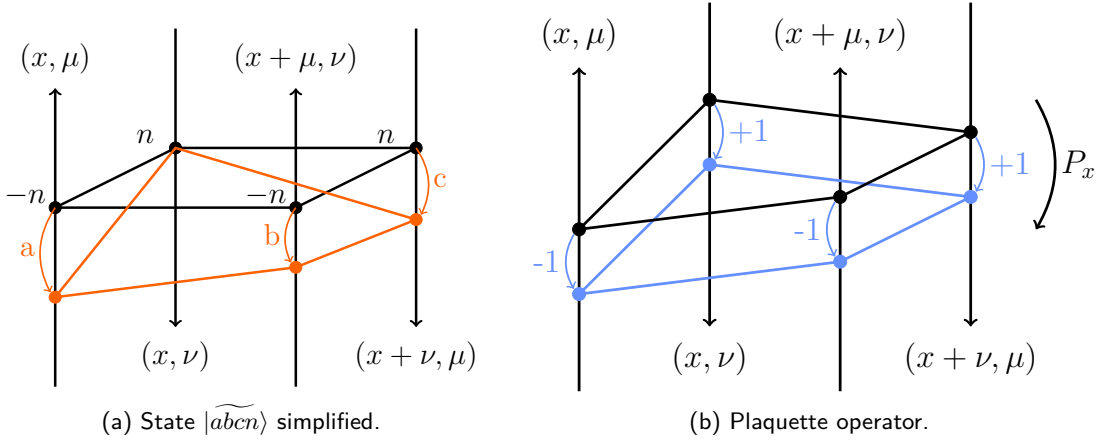
\begin{figure*}[ht!]
    \centering
    \begin{subfigure}{.45\textwidth}
      \centering
      \Large\resizebox{\textwidth}{!}{\renewcommand{\cba}{cborange}
\begin{tikzpicture}

\draw[color=black, very thick, ->] (0,0) -- (0,5);
\draw[color=black, very thick, <-] (2,1) -- (2,6);
\draw[color=black, very thick, ->] (4,0) -- (4,5);
\draw[color=black, very thick, <-] (6,1) -- (6,6);
\node at (0,5.5) {$(x,\mu)$};
\node at (2,0.5) {$(x,\nu)$};
\node at (4,5.5) {$(x+\mu,\nu)$};
\node at (6,0.5) {$(x+\nu,\mu)$};

\filldraw[color=black, fill=black] (0,3) circle (0.1);
\filldraw[color=black, fill=black] (2,4) circle (0.1);
\filldraw[color=black, fill=black] (4,3) circle (0.1);
\filldraw[color=black, fill=black] (6,4) circle (0.1);
\draw[color=black, very thick] (0,3) -- (2,4) -- (6,4) -- (4,3) -- (0,3);
\node at (-0.5,3.2) {$-n$};
\node at (1.5,4.2) {$n$};
\node at (3.5,3.2) {$-n$};
\node at (5.5,4.2) {$n$};

\filldraw[color=\cba] (0,1.5) circle (0.1);
\filldraw[color=\cba] (4,2) circle (0.1);
\filldraw[color=\cba] (6,2.8) circle (0.1);
\draw[color=\cba, very thick] (0,1.5) -- (2,4) -- (6,2.8) -- (4,2) -- (0,1.5);
\draw[->, thick, color=\cba] (0,3) to[bend right] node[midway,left,inner sep=2pt, color=\cba] {a} (-0.08,1.58) ;
\draw[->, thick, color=\cba] (4,3) to[bend right] node[midway,left,inner sep=2pt, color=\cba] {b} (3.92,2.08) ;
\draw[->, thick, color=\cba] (6,4) to[bend left] node[midway,right,inner sep=2pt, color=\cba] {c} (6.08,2.88) ;

\end{tikzpicture}}
      \caption{State $\ket{\widetilde{abcn}}$ simplified.}
      \label{fig:sub-plaquettestate}
    \end{subfigure}
    \begin{subfigure}{.53\textwidth}
      \centering
      \Large\resizebox{\textwidth}{!}{\renewcommand{\cbb}{black}
\renewcommand{\cba}{cbblue}
\begin{tikzpicture}

\draw[color=black, very thick, ->] (0,0) -- (0,5);
\draw[color=black, very thick, <-] (2,1) -- (2,6);
\draw[color=black, very thick, ->] (4,0) -- (4,5);
\draw[color=black, very thick, <-] (6,1) -- (6,6);
\node at (0,5.5) {$(x,\mu)$};
\node at (2,0.5) {$(x,\nu)$};
\node at (4,5.5) {$(x+\mu,\nu)$};
\node at (6,0.5) {$(x+\nu,\mu)$};

\filldraw[color=\cba] (0,1.5) circle (0.1);
\filldraw[color=\cba] (2,3.5) circle (0.1);
\filldraw[color=\cba] (4,2) circle (0.1);
\filldraw[color=\cba] (6,3) circle (0.1);
\draw[color=\cba, very thick] (0,1.5) -- (2,3.5) -- (6,3) -- (4,2) -- (0,1.5);

\filldraw[color=\cbb] (0,2.5) circle (0.1);
\filldraw[color=\cbb] (2,4.5) circle (0.1);
\filldraw[color=\cbb] (4,3) circle (0.1);
\filldraw[color=\cbb] (6,4) circle (0.1);
\draw[color=\cbb, very thick] (0,2.5) -- (2,4.5) -- (6,4) -- (4,3) -- (0,2.5);
\draw[<-, thick, color=\cba] (-.08,1.58) to[bend left] node[midway,left,inner sep=2pt, color=\cba] {-1} (-0.08,2.42) ;
\draw[<-, thick, color=\cba] (2.08,3.58) to[bend right] node[midway,right,inner sep=2pt, color=\cba] {+1} (2.08,4.42) ;
\draw[<-, thick, color=\cba] (3.92,2.08) to[bend left] node[midway,left,inner sep=2pt, color=\cba] {-1} (3.92,2.92) ;
\draw[<-, thick, color=\cba] (6.08,3.08) to[bend right] node[midway,right,inner sep=2pt, color=\cba] {+1} (6.08,3.92) ;

\draw[<-, very thick, color=\cbb] (7,2.5) to[bend right] node[midway,right,inner sep=2pt, color=\cbb] {$P_x$} (7,4.5) ;
\end{tikzpicture}}
      \caption{Plaquette operator.}
      \label{fig:sub-plaquetteoperator}
    \end{subfigure}
    \caption{Plaquette states and operator represented in the subspace $\mathbb{Z}^4$ containing four gauge field values.}
    \label{fig:weirdket}
\end{figure*}

The state and plaquette operator are illustrated in Fig. \ref{fig:weirdket}.

\paragraph*{Gauge field truncation.}\label{sec:gaugetruncation} If truncating the gauge field to $\mathcal{H}_k$, one needs $\ket{\widetilde{abc,k,o}} = \ket{\widetilde{abc,0,o}}$ so that any sign discrepancy in Eq. \eqref{eq:plaquetteasshift} is avoided. This paragraph is just to overcome this technicality. In terms of phase the condition is that $\psi_{abcko} = \psi_{abc0o}=0$, i.e. that $\psi_{abcko}=\sum_{0\leq n<k}\varphi_{abcno}$ be even. Each term $\varphi_{abcno}$ is actually a sum $\varphi_{abcn} + \varphi_o$, where $\varphi_{abcn}$ depends on the four links under transformation, whereas $\varphi_o$ depends on the other gauge fields and qubits at the four sites, which are not modified by the plaquette operator. Both of these depend on the JW order chosen. Since $\varphi_o$ is constant when acting only using the plaquette term, for an even $k$ its contribution to the sum $\sum_{0\leq n < k} \varphi_o$ is even. Additionally, for any $n$, we have that $\varphi_{abc,n}=\varphi_{abc,n+2}$ as all the parities are equal. As a consequence $\varphi_{abc,n}+\varphi_{abc,n+1}+\varphi_{abc,n+2}+\varphi_{abc,n+3}$ is even. It follows that for $k$ a multiple of $4$, the sum $\sum_{0\leq n < k} \varphi_{abcn}$ is even. Hence, for $k=4q$, with $q$ an integer, the truncation is valid. 

Relying on a specific JW order actually allows for any even $k$. Indeed, we can enforce that $\varphi_{abcn}$ always be equal to $0$: this is the case when $(x:-\eta)$ and $x:\eta$ are both inferior (or both superior) to $(x:-\zeta)$ and $(x:\zeta)$ for $\eta$ and $\zeta$ two distinct directions. In this case, the four corner operators that form a plaquette defined in Eq. \eqref{eq:cornerop} will induce a phase $(-1)^{\varphi_{abcn}}$ through the following $Z$ operators (disregarding the part which contributes to $\varphi_{o}$): 
$$Z_{x:\eta} Z_{x+\eta:-\eta} Z_{x+\eta+\zeta:-\eta} Z_{x+\zeta:\eta}.$$
Notice that each link appear twice, hence they do not induce any phase, i.e. we have that $P\ket{\widetilde{abcno}}=(-1)^{\varphi_o} \ket{\widetilde{abc,n+1,o}}$. Thus, the truncation is well-defined for $k$ even for that specific JW orders.

\subsubsection{Derivation through diagonalization}
The first construction moves from the electric to the magnetic basis \cite{Haase2021ARE}. In other words, it works by computing the eigenvectors and eigenvalues of the plaquette operator, so that taking the exponential of the operator simply amounts to exponentiating the eigenvalues. 

A single plaquette local evolution can be seen as a shift operator upon plaquette states as illustrated Fig. \ref{fig:sub-plaquetteoperator}. That shift amounts to a phase in the Fourier basis, therefore the Fourier transform diagonalizes the plaquette. To define this Fourier transform, the state space of the gauge field needs be truncated to $\mathbb{Z}_k$, for instance according to Remark \ref{rem:truncationE}, and in accordance with the restriction of Subsec. \eqref{sec:magcontrib}. Then the eigenvectors of a plaquette local evolution are 
\begin{equation}
    \ket{abcp}_\square = \frac{1}{\sqrt{k}} \sum_{n=0}^{k-1} e^{2\pi i n p/k} \ket{\widetilde{abc,n}}
\end{equation}
for $a,b,c$ and $p$ integers in $0,..., k-1$. These are the Fourier transform of the $\ket{\widetilde{abc,n}}$ states, that can be obtained through a quantum Fourier transform (FT) whose circuit representation is well-known \cite{ShorQFT}. The corresponding eigenvalues are 
\begin{equation}
    \lambda_{p} = e^{2\pi i p/k}.
\end{equation}
The eigenvectors of the hermitian conjugate of the plaquette operator are the same but with eigenvalues $\lambda_{-p}$. For the sum of the plaquette and its hermitian conjugate, the eigenvalues are:
\begin{equation}
    e^{2\pi i p/k} + e^{-2\pi i p/k} = 2 \cos(2\pi p/k).
\end{equation}


Having found an eigenbasis of the plaquette term, it is now easy to exponentiate it. Doing so one gets the eigenvalue equation:
\begin{widetext}\begin{align}
    \exp \left( i\epsilon^2 \frac{g_M^2}{2} (P + P^\dagger)\right)\ket{abcp}_\square
        &= \exp \left(i\epsilon^2 g_M^2 \cos(2\pi p/k)\right) \ket{abcp}_\square.
\end{align}\end{widetext}
This defines a diagonal operator $\text{Diag}$ that contains the eigenvalues $\exp(i\epsilon^2 g_M^2 \cos(2\pi p/k))$. The magnetic evolution for the QCA can thus be written as:
\begin{equation}\label{eq:qcamag1first}
    \mathbf{D_M^{(0)}} = \bigotimes_x \ (\text{FT}^\dagger ~\text{Diag}~ \text{FT})
\end{equation}
where $\text{FT}$ is the quantum Fourier transform. Hence, Eq. \eqref{eq:qcamag1first} thus defines a circuit of quantum gates for the magnetic term. In this construction, no approximation has been done, hence the limit is exactly the one given in Eq. \eqref{eq:qcamag1limit}.

\subsubsection{Quantum walk-like derivation}
The second construction uses a reformulation of the plaquette operators to make it look like a quantum walk. Here we sometime abuse notations and write $\ket{\widetilde{abcn}}$ or even just $\ket{\widetilde{n}}$ to talk about a state $\ket{\widetilde{abcno}}$. Indeed, the plaquette term can be divided into two operators, one which acts as $\ket{\widetilde{2n}}\bra{\widetilde{2n+1}}+\ket{\widetilde{2n+1}}\bra{\widetilde{2n}}$ and the other as $\ket{\widetilde{2n+1}}\bra{\widetilde{2n+2}} + \ket{\widetilde{2n+2}}\bra{\widetilde{2n+1}}$---i.e. the two operators act as shifted swaps between pairs of sites. Let $\widetilde{P}$ and $\widetilde{Q}$ denote those operators:
\begin{widetext}\begin{align}
    \widetilde{P} &= \sum_{abc \in \mathbb{Z}^3} \sum_{n \in 2\mathbb{Z}} \ket{\widetilde{abc,n+1}}\bra{\widetilde{abc,n}} + \ket{\widetilde{abc,n}}\bra{\widetilde{abc,n+1}} \\
    \widetilde{Q} &= \sum_{abc \in \mathbb{Z}^3} \sum_{n \in 2\mathbb{Z}+1} \ket{\widetilde{abc,n+1}}\bra{\widetilde{abc,n}} + \ket{\widetilde{abc,n}}\bra{\widetilde{abc,n+1}}.
\end{align}\end{widetext}
We have that 
\begin{equation}
    \widetilde{P} + \widetilde{Q} = P + P^\dagger.
\end{equation}

$\widetilde{P}$ and $\widetilde{Q}$ are hermitian, unitary and their eigenvectors are
\begin{align}
    \ket{+_n} &= \frac{1}{\sqrt{2}}(\ket{\widetilde{n}} + \ket{\widetilde{n+1}}) \\
    \ket{-_n} &= \frac{1}{\sqrt{2}}(\ket{\widetilde{n}} - \ket{\widetilde{n+1}}) 
\end{align}
with $n$ even for $\widetilde{P}$ and odd for $\widetilde{Q}$. The corresponding eigenvalues are plus and minus one.

Now taking the exponential of $\widetilde{P}$, we have, with $n$ even:
\begin{align}
    \exp\left( i \epsilon^2 \frac{g_M^2}{\sqrt{2}} \widetilde{P} \right) \ket{+_n} &= \exp \left(i \epsilon^2 \frac{g_M^2}{\sqrt{2}}\right) \ket{+_n} \\
    \exp\left( i \epsilon^2 \frac{g_M^2}{\sqrt{2}} \widetilde{P} \right) \ket{-_n} &= \exp \left(-i \epsilon^2 \frac{g_M^2}{\sqrt{2}}\right) \ket{-_n}
\end{align}
\begin{widetext}\begin{align}
    \exp\left( i \epsilon^2 \frac{g_M^2}{2} \widetilde{P} \right) \ket{\widetilde{n}} &=\frac{1}{\sqrt{2}} \left( \exp \left(i \epsilon^2 \frac{g_M^2}{2}\right) \ket{+_n} + \exp \left(-i \epsilon^2 \frac{g_M^2}{2}\right) \ket{-_n} \right) \\
    &=\cos(\epsilon^2 \frac{g_M^2}{2}) \ket{\widetilde{n}} + i \sin(\epsilon^2 \frac{g_M^2}{2})\ket{\widetilde{n+1}}\\
    \exp\left( i \epsilon^2 \frac{g_M^2}{2} \widetilde{P} \right) \ket{\widetilde{n+1}}
    &= i \sin(\epsilon^2 \frac{g_M^2}{2})\ket{\widetilde{n}} + \cos(\epsilon^2 \frac{g_M^2}{2}) \ket{\widetilde{n+1}}.
\end{align}

These equations are identical for $\exp\left( i \epsilon^2 \frac{g_M^2}{2} \widetilde{Q} \right)$ when taking $n$ odd.

This is reminiscent of a one dimensional quantum walk with coin 
\begin{equation}
    \begin{pmatrix}
        \cos(\theta)    &   i \sin(\theta)\\
        i \sin(\theta)  &   \cos(\theta)
    \end{pmatrix}.
\end{equation}
Such QW is homogeneous in the `tilde' basis, but at first glance, it seems inhomogeneous in the canonical basis because of the operator
\begin{equation}
    \ket{\widetilde{abc,n+1,o}}\bra{\widetilde{abc,n,o}} = (-1)^{\varphi_{abcno}} \ket{\underline{abc,n+1,o}}\bra{\underline{abc,n,o}}
\end{equation} 
which induces a phase $\varphi_{abcno}$, dependent upon $n$, in the canonical basis. However, as explained in Subsec. \ref{sec:gaugetruncation}, this phase can be made independent of $n$ through a specific choice of JW order, recovering a homogeneous QW even in the canonical basis.\\
Consider
\begin{align*}
\mathbf{D}_{\widetilde{P}}&=\bigotimes_x \exp\left( i \epsilon^2 \frac{g_M^2}{2} \widetilde{P} \right)\\
\mathbf{D}_{\widetilde{Q}}&=\bigotimes_x \exp\left( i \epsilon^2 \frac{g_M^2}{2} \widetilde{Q} \right).
\end{align*}
A way to implement the plaquette term is to apply those successively in one time step:
\begin{equation}\label{eq:qcamag2}
    \mathbf{D_M^{(1)}} \ket{\widetilde{n}} = \mathbf{D_{\widetilde{Q}} D_{\widetilde{P}}}.
\end{equation}

Indeed, taking the limit when $\epsilon$ goes to zero after applying both the exponentials of $\widetilde{P}$ and $\widetilde{Q}$ to $\ket{\widetilde{n}}$ (where $n$ is even) gives:
\begin{align}
    \exp\left( i \epsilon^2 \frac{g_M^2}{2} \widetilde{Q} \right)\exp\left( i \epsilon^2 \frac{g_M^2}{2} \widetilde{P} \right) \ket{\widetilde{n}} &= \exp\left( i \epsilon^2 \frac{g_M^2}{2} \widetilde{Q} \right) \left( \ket{\widetilde{n}} + i \epsilon^2 \frac{g_M^2}{2}\ket{\widetilde{n+1}} + O\left(\epsilon^4\right)\right) \\
    &=\ket{\widetilde{n}} + i\epsilon^2 \frac{g_M^2}{2} \left( \ket{\widetilde{n-1}} + \ket{\widetilde{n+1}} \right) + O\left(\epsilon^4\right)\\
    &=\left(I + i\epsilon^2 \frac{g_M^2}{2} \left( P^\dagger + P \right)\right)\ket{\widetilde{n}} + O\left(\epsilon^4\right)
\end{align}\end{widetext}

For $n$ odd, the exponential of $\widetilde{P}$ would yield the state $\ket{\widetilde{n-1}}$ instead of $\ket{\widetilde{n+1}}$ (in the second line of the equation) and the exponential of $\widetilde{Q}$ the state $\ket{\widetilde{n+1}}$ instead of $\ket{\widetilde{n-1}}$ (in the third line). Therefore, the same limit would be reached.

In $\mathbf{D_M^{(1)}}$ this is done at every plaquette, matching Eq. \eqref{eq:qcamag1limit}. Thus, both quantum gate implementations agree in the limit as $\mathbf{D_M^{(0)}}=\mathbf{D_M^{(1)}}+O(\epsilon^4)$. 

\paragraph*{Gauge invariance.} 

Gauge invariance of the plaquette term has been verified in Sec. \ref{sec:gioperators}. Its exponentiation is a linear combination of them, thus still gauge invariant. 


\subsection{Complete dynamics}

Combining the fermionic dynamics, the electric and magnetic contribution, one obtains the following complete dynamics for the $2+1$ QED QCA:
\begin{equation}\label{eq:fullqca2d}
    \text{\bf{QCA}} = \mathbf{D_M D_E D_F}
\end{equation}
where $\mathbf{D_F}$ (fermionic term) refers to Eq. \eqref{eq:qcafermion}, $\mathbf{D_E}$ (electric term) refers to Eq. \eqref{eq:qcaelectric} and $\mathbf{D_M}$ (magnetic term) refers to Eq. \eqref{eq:qcamag1} or \eqref{eq:qcamag2} depending on the quantum gate implementation chosen.

\section{3+1 Quantum Cellular Automaton}
\label{sec:evolution3d}

This section extends the previous $2+1$ QED QCA construction, to reach a $3+1$ QED QCA. In $3+1$ dimensions, the Dirac Eq. is a PDE on a wave function having four complex amplitudes, i.e. there are four fermionic modes instead of two. As for the gauge field, no additional degree of freedom is required, but the third dimension needs to be taken into account when considering the electronic and plaquette terms.

In the following, to avoid mixing notations, every operator referring to the $3+1$ case will be overlined, e.g. the on-site operator modelling the mass in the $3+1$ Dirac QW will be denoted $\overline{C}_\epsilon$. Again the operators of the multi-particle sector QCA are in bold, e.g. $\overline{\mathbf{C}}_\epsilon$.

\subsection{Fermionic dynamics}\label{sec:3dfermdyn}
The procedure is the same here as in the $2+1$ case: (i) define the $3+1$ Dirac QW, (ii) extend each on-site operator using the Heisenberg picture so that it acts on the full state space and not just in the one-particle sector---cf. Appendix \ref{appendix:qwtoqca}.

Let $\gamma_j$ be the following generalized Pauli matrix:
\begin{align}
    \gamma_0 &= Y \otimes I \\
    \gamma_1 &= \gamma_\nu =  Z \otimes X \\
    \gamma_2 &= \gamma_\mu =  Z \otimes Y\\
    \gamma_3 &= \gamma_\kappa = Z \otimes Z.
\end{align}
They respect the anti-commutation relation
\begin{equation}
    \{ \gamma_j, \gamma_k \} = 2\delta_{j,k} I_4
\end{equation}
with $\delta_{j,k}$ the Kronecker delta.

The state space for the Dirac QW is that of 4 qubits restricted to the one-particle sector, i.e. there are $5$ possible states: $\ket{0000}$, $\ket{0001}$, $\ket{0010}$, $\ket{0100}$ and $\ket{1000}$. As in the $2+1$ case, state $\ket{0000}$ is mapped to itself because of number conservation. Hence, every on-site operator is of the form $1\oplus U$ where $U$ acts on the remaining four dimensional subspace. The $3+1$ Dirac QW has the same structure as that in $2+1$: transport sub-steps in each direction, each of these being surrounded by a basis change, and lastly a mass sub-step \cite{di2012discrete,Arrighi2014TheDE,marquez2017fermion}:
\begin{widetext}\begin{align}
    \overline{\text{QW}} = &
        \left(\bigotimes_{x} \overline{C}_\epsilon \right)\left(\bigotimes_{x} \overline{B} \right)\\ 
        &  \left(\bigotimes_{x} \overline{S}_2~~\bigotimes_{(x,2)(x,3)(x+\mu,0)(x+\mu,1)} \overline{T}_\mu~~\bigotimes_{x} \overline{S}_1\right) \left(\bigotimes_{x} \overline{B} \right) \\
        &\left(\bigotimes_{x} \overline{S}_2~~\bigotimes_{(x,2)(x,3)(x+\nu,0)(x+\nu,1)} \overline{T}_\nu ~~\bigotimes_{x} \overline{S}_1\right) \left(\bigotimes_{x} \overline{B} \right) \\
        &\left(\bigotimes_{x} \overline{S}_2~~\bigotimes_{(x,2)(x,3)(x+\kappa,0)(x+\kappa,1)} \overline{T}_\kappa~~\bigotimes_{x} \overline{S}_1\right) 
        \label{eq:3dqw}
\end{align}
where 
\begin{itemize}
\item $\overline{S}_1$ swaps qubits $(x,0)$ with $(x,1)$, and then $(x,1)$ with $(x,2)$.  
\item $\overline{S}_2$ swaps qubits $(x,1)$ with $(x,2)$, and then $(x,2)$ with $(x,3)$, 
\item $\overline{T}_{\eta}$ swaps qubits $(x,2)$ and $(x,3)$ with $(x+\eta,0)$ and $(x+\eta,1)$ 
\end{itemize}
so that
\begin{equation}
\bigotimes_{x} \overline{S}_2~ \bigotimes_{(x,2)(x,3)(x+\eta,0)(x+\eta,1)} \overline{T}_{\eta} ~\bigotimes_{x} \overline{S}_1 = 1\oplus e^{\epsilon (Z \otimes Z) \partial_\eta}\label{eq:3Dtransport}
\end{equation}
transporting the first and last qubits in the positive direction and the middle two qubits in the opposite one.

The operators $\overline{B}$ is defined so that, \PA{as acting on some $\ket{\psi}^{(x,3)(x,2)(x,1)(x,0)}$}:
\begin{equation}
    \overline{B}^\dagger (1\oplus \gamma_\eta) \overline{B} = 1\oplus \gamma_{(\eta+1\mod 3)+1},\label{eq:Bincrement}
\end{equation}\end{widetext}
i.e. cycling from $\gamma_\kappa$ to $\gamma_\mu$, from $\gamma_\mu$ to $\gamma_\nu$, and back.
One possible choice is given in \cite{ArrighiTetrahedra}. First, one notices that $R_{\sigma_z}(\theta)R_{\sigma_x}(\theta)$, with $\theta=\pi/2$, maps the Bloch vector of the Pauli matrix $\sigma_{\eta}$ into that of $\sigma_{(\eta+1\mod 3)+1}$ with $\mu=1,2,3$, i.e:
\begin{equation}
     \sigma_{(\eta+1\mod 3)+1} = R_{\sigma_z}(\theta)R_{\sigma_x}(\theta)\sigma_{\eta} R^{\dagger}_{\sigma_x}(\theta)R^{\dagger}_{\sigma_z}(\theta)
\end{equation}
since $(R_{\sigma_z}(\theta)R_{\sigma_x}(\theta))^3=-\mathbb{I}$, and we want it to be the identity, we define 
\begin{align*}
B=e^{-\mathrm{i}\frac{\pi}{3}}R^\dagger_{\sigma_x}(\theta)R^\dagger_{\sigma_z}(\theta)=\frac{1}{\sqrt{2}}\begin{pmatrix}
e^{-i\pi/12} & i e^{-i7\pi/12}\\
i e^{-i\pi/12} & e^{-i7\pi/12}
\end{pmatrix}.
\end{align*}
Lastly we carry this though to the $\gamma_\eta$ by letting
\begin{align*}
\overline{B}=1\oplus(I\otimes B).
\end{align*}

The whole point of Eqs. \eqref{eq:3dqw} and \eqref{eq:Bincrement} is to approach the $\epsilon$ time evolution of the $3+1$-Dirac Hamiltonian $D=m\gamma_0+i\gamma_\mu\partial_\mu+i\gamma_\nu\partial_\nu+i\gamma_\kappa \partial_\kappa$ as follows:
\begin{align*}
&\exp(\epsilon(-im\gamma_0+\gamma_\mu\partial_\mu+\gamma_\nu\partial_\nu+\gamma_\kappa \partial_\kappa))\\ 
&\approx \exp(-im\epsilon\gamma_0)\exp(\gamma_\mu\epsilon\partial_\mu)\exp(\gamma_\nu\epsilon\partial_\nu)\exp(\gamma_\kappa\epsilon\partial_\kappa)\\
&= \exp(-im\epsilon\gamma_0){(I\otimes B)^2}^\dagger\exp(\gamma_\kappa\epsilon\partial_\mu)(I\otimes B)^2(I\otimes B)^\dagger\exp(\gamma_\kappa\epsilon\partial_\nu)(I\otimes B)\exp(\gamma_\kappa\epsilon\partial_\kappa)\\
&= \exp(-im\epsilon\gamma_0)(I\otimes B)\exp(\gamma_\kappa\epsilon\partial_\mu)(I\otimes B)\exp(\gamma_\kappa\epsilon\partial_\nu)(I\otimes B)\exp(\gamma_\kappa\epsilon\partial_\kappa).
\end{align*}
Adding the trivial zero particle sector indeed gives:
\begin{align*}
(1\oplus\exp(-\epsilon im\epsilon\gamma_0))
\overline{B}
(1\oplus e^{\epsilon (Z \otimes Z) \partial_\mu)}
\overline{B}
(1\oplus e^{\epsilon (Z \otimes Z) \partial_\nu)}
\overline{B}
(1\oplus e^{\epsilon (Z \otimes Z) \partial_\kappa}).
\end{align*}
We see that the on-site operator for the mass in $3+1$ dimensions is quite similar to that of the $2+1$ dimensional case. We have, \PA{as acting on some $\ket{\psi}^{(x,3)(x,2)(x,1)(x,0)}$}:
\begin{align}
    \overline{C}_\epsilon &= 1 \oplus e^{-im\epsilon\gamma_0}  \\
        &= 1 \oplus (c I_4 -i s\gamma_0) \\
        &= 1 \oplus \begin{pmatrix}
            c & 0 & -s & 0 \\
            0 & c & 0 & -s \\
            s & 0 & c & 0 \\
            0 & s & 0 & c
        \end{pmatrix} \\
        &= 1 \oplus  (1\oplus X \oplus 1)(C_\epsilon\oplus C_\epsilon)(1\oplus X \oplus 1) \label{eq:Cas2qubits}
\end{align}
with $c=\cos(m\epsilon)$ and $s=\sin(m\epsilon)$.

\medskip

Next we must convert this $3+1$ Dirac QW into a $3+1$ Dirac QCA. The derivation of the operators making up the $3+1$ Dirac QCA requires us to extend the on-site operators any configuration for the $4$ qubits, hence possibly more than two fermions crossing. In general this could be complicated to analyse. Fortunately, all of the above $4$-qubit on-site operators can be reexpressed as products of $2$-qubit on-site gates, with the $2$ qubits being adjacent in the JW order. Thanks to this, the $2+1$ methodology and results readily apply, see Appendix \ref{appendix:qwtoqca} for details.

\begin{figure}[ht!]
    \centering
    \large\resizebox{!}{8em}{\renewcommand{\cba}{cbpurple}
\renewcommand{\cbb}{cbyellow}
\renewcommand{\cbc}{cborange}
\begin{tikzpicture}

\newcommand{\state}[2]{\filldraw[color=black, fill=black] (#1,#2) circle (0.1);}
\newcommand{\states}[2]{
    \begin{scope}[shift={(#1,#2)}] 
        \state{0}{0};
        \state{1}{0};
        \state{2}{0};
        \state{3}{0};
    \end{scope}
}
\newcommand{\naming}[6]{
    \begin{scope}[shift={(#1,#2)}] 
        \node at (0,0) {#3};
        \node at (1,0) {#4};
        \node at (2,0) {#5};
        \node at (3,0) {#6};
    \end{scope}
}
\newcommand{\gate}[3]{
\begin{scope}[shift={(#1,#2)}] 
    \draw[thick] (0,0) -- (1,1) (1,0) -- (0,1);
    \filldraw[color=black, fill=white] (0.2,0.2) rectangle (0.8,0.8);
    \node at (.5,.5) {#3};
\end{scope}}
\newcommand{\swapper}[2]{
\begin{scope}[shift={(#1,#2)}] 
    \gate{1}{0}{$\mathbf{S}$}
    \gate{0}{1}{$\mathbf{C}_\epsilon$}
    \gate{2}{1}{$\mathbf{C}_\epsilon$}
    \gate{1}{2}{$\mathbf{S}$}
    \draw[thick] (0,0) -- (0,1);
    \draw[thick] (0,2) -- (0,3);
    \draw[thick] (3,0) -- (3,1);
    \draw[thick] (3,2) -- (3,3);
    \draw[very thick, color=\cbb] (-.2,0.15) rectangle (3.2,2.85);
\end{scope}}

\states{0}{0}
\naming{0}{-.5}{0}{1}{2}{3}

\swapper{0}{0}
\node[color=\cbb] at (4,1.5) {$\overline{\mathbf{C}}_\epsilon$};

\begin{scope}[shift={(0,3)}] 
    \states{0}{0}
    \naming{0}{.5}{0}{1}{2}{3}
\end{scope}

\end{tikzpicture}}
    \caption{Representation of the mass term of $3+1$ QED QCA.}
    \label{fig:mass}
\end{figure}
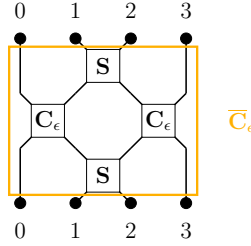

\paragraph{The mass sub-step} of the $3+1$ Dirac QCA is reached by: 1/ Seeking to represent $\overline{C}_\epsilon$ as a circuit of one-particle-sector $2$-qubit gates, as done in Eq. \eqref{eq:Cas2qubits}. 2/ Replacing each $2$-qubit gate of this circuit by its multi-particle sector extension, and the direct sums by tensor products, so as to obtain the $3+1$ Dirac QCA on-site operator $\overline{\mathbf{C}}_\epsilon$. 3/ Repeating $\overline{\mathbf{C}}_\epsilon$ across space.\\ 
This procedure is justified in Appendix \ref{appendix:qwtoqca}, essentially on the basis that the QCA should act as the QW in the one-particle sector. 
We get, \PA{as acting on some $\ket{\psi}^{(x,3)(x,2)(x,1)(x,0)}$}:
\begin{equation}
    \overline{\mathbf{C}}_\epsilon = (I\otimes \mathbf{S}\otimes I)(\mathbf{C}_\epsilon\otimes \mathbf{C}_\epsilon) (I\otimes \mathbf{S}\otimes I)
\end{equation}
where $\mathbf{S}$ is the swap defined in Eq. \eqref{eq:swap} and $\mathbf{C}_\epsilon$ is the 2+1D mass term defined in Eq. \eqref{eq:2dmass}. This is represented in Fig. \ref{fig:mass}.

\paragraph{The basis changes substep} of the $3+1$ Dirac QCA is obtained by following the same procedure. 1/ The $3+1$ Dirac QW can again be expressed as a direct sum of two qubit gates 
\begin{equation}
\overline{B}=1\oplus(B\oplus B). \label{eq:ovB}
\end{equation}
2/ We build the QCA version of $B$ as prescribed by \eqref{eq:woperatorqcafromqw}, yielding, as acting on some $\ket{\psi}^{(x,1)(x,0)}$,
\begin{equation}
    \mathbf{B} = \begin{pmatrix}
1& 0& 0 &0\\
0& \frac{1}{\sqrt{2}}e^{-i\pi/12} &  \frac{i}{\sqrt{2}}e^{-i7\pi/12}&0\\
0& \frac{i}{\sqrt{2}} e^{-i\pi/12} & \frac{1}{\sqrt{2}} e^{-i7\pi/12}&0\\
0& 0& 0& e^{-i2\pi/3}
\end{pmatrix}.\label{eq:2DqcaB}
\end{equation}
and replace $B$ by $\mathbf{B}$ and $\oplus$ by $\otimes$ in \eqref{eq:ovB} to get:
\begin{equation}
\overline{\mathbf{B}}=\mathbf{B}\otimes \mathbf{B} \label{eq:ovbfB}
\end{equation}
3/ This is what needs be applied across space to make up the basis changes substep of the $3+1$ Dirac QCA.

\begin{figure*}[ht!]
    \centering
    \resizebox{\textwidth}{!}{\newcommand{\colora}{\cba}
\newcommand{\colorb}{\cbb}
\newcommand{\colorc}{\cbc}
\begin{tikzpicture}

\newcommand{\state}[2]{\filldraw[color=black, fill=black] (#1,#2) circle (0.1);}
\newcommand{\states}[2]{
    \begin{scope}[shift={(#1,#2)}] 
        \state{0}{0};
        \state{1}{0};
        \state{2}{0};
        \state{3}{0};
    \end{scope}
}
\newcommand{\naming}[6]{
    \begin{scope}[shift={(#1,#2)}] 
        \node at (0,0) {#3};
        \node at (1,0) {#4};
        \node at (2,0) {#5};
        \node at (3,0) {#6};
    \end{scope}
}
\newcommand{\gate}[3]{
\begin{scope}[shift={(#1,#2)}] 
    \draw[thick] (0,0) -- (1,1) (1,0) -- (0,1);
    \filldraw[color=black, fill=white] (0.2,0.2) rectangle (0.8,0.8);
    \node at (.5,.5) {#3};
\end{scope}}

\newcommand{\swapper}[2]{
\begin{scope}[shift={(#1,#2)}] 
    \gate{1}{0}{$\mathbf{S}$}
    \gate{0}{1}{$\mathbf{S}$}
    \gate{2}{1}{$\mathbf{S}$}
    \gate{1}{2}{$\mathbf{S}$}
    \draw[thick] (0,0) -- (0,1);
    \draw[thick] (0,2) -- (0,3);
    \draw[thick] (3,0) -- (3,1);
    \draw[thick] (3,2) -- (3,3);
    \draw[color=\colora, very thick] (-.2,0.15) rectangle (3.2,2.85);
\end{scope}}

\newcommand{\swapperone}[2]{
\begin{scope}[shift={(#1,#2)}] 
    \gate{0}{0}{$\mathbf{S}$}
    \gate{1}{1}{$\mathbf{S}$}
    \draw[thick] (0,1) -- (0,2);
    \draw[thick] (2,0) -- (2,1);
    \draw[thick] (3,0) -- (3,2);
    \draw[color=\colora, very thick] (-.2,0.15) rectangle (3.2,1.85);
\end{scope}}

\newcommand{\swappertwo}[2]{
\begin{scope}[shift={(#1,#2)}] 
    \gate{2}{1}{$\mathbf{S}$}
    \gate{1}{0}{$\mathbf{S}$}
    \draw[thick] (3,0) -- (3,1);
    \draw[thick] (1,1) -- (1,2);
    \draw[thick] (0,0) -- (0,2);
    \draw[color=\colora, very thick] (-.2,0.15) rectangle (3.2,1.85);
\end{scope}}

\newcommand{\transport}[2]{
\begin{scope}[shift={(#1,#2)}] 
    \gate{1}{0}{$\mathbf{T}_{\eta}$}
    \gate{0}{1}{$\mathbf{S}$}
    \gate{2}{1}{$\mathbf{S}$}
    \gate{1}{2}{$\mathbf{T}_{\eta}$}
    \draw[thick] (0,0) -- (0,1)
            (0,2) -- (0,3)
            (3,0) -- (3,1)
            (3,2) -- (3,3);
    \draw[thick] (-1,-1) -- (0,0)
            (0,-1) -- (1,0)
            (0,3) -- (-1,4)
            (1,3) -- (0,4)
            (4,-1) -- (3,0)
            (3,-1) -- (2,0)
            (3,3) -- (4,4)
            (2,3) -- (3,4);
        
    \draw[color=\colorc, very thick] (-.2,0.15) rectangle (3.2,2.85);
\end{scope}}

\states{-6}{0}
\naming{-6}{-.5}{0}{1}{2}{3}
\states{0}{0}
\naming{0}{-.5}{0}{1}{2}{3}
\states{6}{0}
\naming{6}{-.5}{0}{1}{2}{3}

\swapperone{-6}{0}
\swapperone{0}{0}
\swapperone{6}{0}
\node[color=\colora] at (8,2.5) {$\overline{\mathbf{S}}_1$};

\transport{-3}{3}
\transport{3}{3}
\node[color=\colorc] at (8,4.5) {$\overline{\mathbf{T}}_{\eta}$};

\swappertwo{-6}{7}
\swappertwo{0}{7}
\swappertwo{6}{7}
\node[color=\colora] at (8,6.5) {$\overline{\mathbf{S}}_2$};

\begin{scope}[shift={(0,9)}] 
    \states{-6}{0}
    \naming{-6}{.5}{0}{1}{2}{3}
    \states{0}{0}
    \naming{0}{.5}{0}{1}{2}{3}
    \states{6}{0}
    \naming{6}{.5}{0}{1}{2}{3}
\end{scope}

\end{tikzpicture}}
    \caption{Representation of the transport}
    \label{fig:transport3}
\end{figure*}
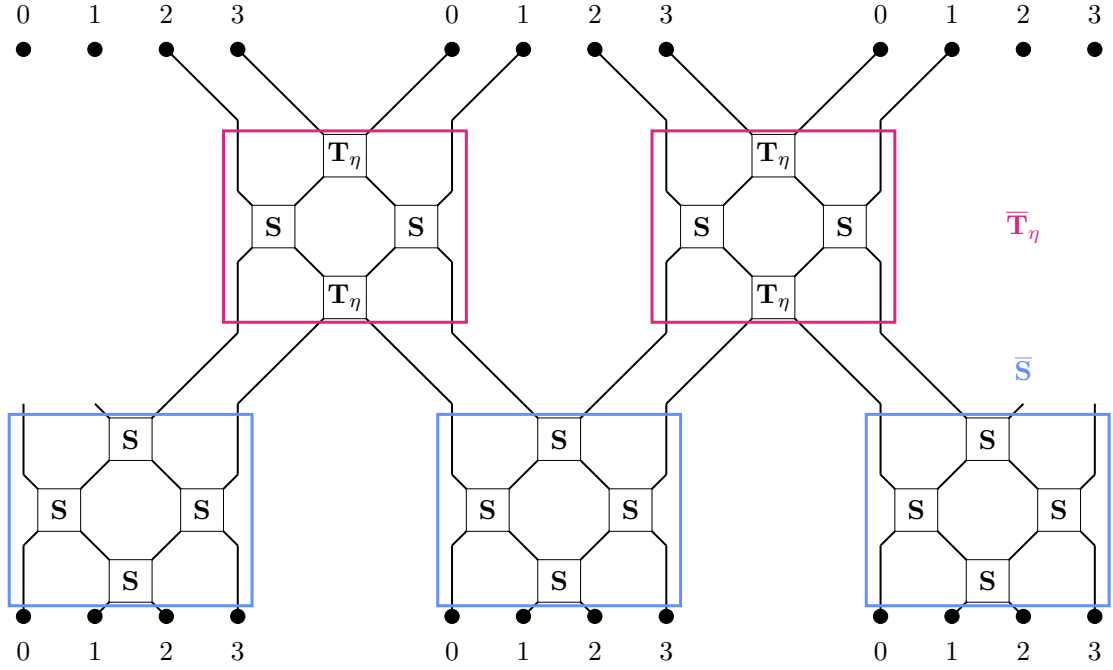

\paragraph{The transport sub-step} of the $3+1$ Dirac QCA is recovered from that of the $3+1$ Dirac QW in a similar fashion. It moves the first and last qubits in the $\eta$ direction while moving the middle two qubits in the direction $-\eta$. 
This is done by first swapping the first two qubits with one another, and then the middle qubits with one another, using a on-site operator $\overline{\mathbf{S}}_1$, second by applying a transport operator which will swap these pairs of qubits with neighbouring ones in the $\eta$ direction using an across-two-sites operator $\overline{\mathbf{T}}_{\eta}$, and finally by first swapping the middle two qubits with one another, and then the last two qubits with one another, using a on-site operator $\overline{\mathbf{S}}_2$.
Again each of these is obtained by decomposing the corresponding $3+1$ Dirac QW on-sites operators into two qubit-gates, and replacing them by their multi-particle sector extensions, as illustrated in Fig. \ref{fig:transport3}.\\
For $\overline{\mathbf{S}}_1$ we get, \PA{as acting on some $\ket{\psi}^{(x,3)(x,2)(x,1)(x,0)}$}:
\begin{align}
    \overline{\mathbf{S}}_1 &= (I\otimes \mathbf{S}\otimes I) (I\otimes I\otimes \mathbf{S})
\end{align}
For $\overline{\mathbf{S}}_2$ we get, \PA{as acting on some $\ket{\psi}^{(x,3)(x,2)(x,1)(x,0)}$}:
\begin{align}
    \overline{\mathbf{S}}_2 &=  ( \mathbf{S}\otimes I\otimes I)(I\otimes \mathbf{S}\otimes I) 
\end{align}
For $\overline{\mathbf{T}}_{\eta}$ we get, \PA{as acting on some $\ket{\psi}^{(x+\eta,1)(x+\eta,0)(x,3)(x,2)}$}:
\begin{align}    
    \overline{\mathbf{T}}_{\eta} &= (I\otimes \mathbf{T}_{\eta} \otimes I) (\mathbf{S}\otimes\mathbf{S})(I\otimes \mathbf{T}_{\eta} \otimes I)
\end{align}
where $\mathbf{S}$ is defined as in Eq. \eqref{eq:swap}, and $\mathbf{T}_{\eta}$ is defined as in \eqref{eq:transport} but with $j=3$.

\paragraph{The full fermionic dynamics} is therefore:
\PA{
\begin{align}
    \mathbf{\overline{D}_F} =  &\left[\left(\bigotimes_x \mathbf{\overline{C}}_\epsilon\right) \left(\bigotimes_x \mathbf{\overline{B}} \right)\right]\\
    &  \left[\left(\bigotimes_x \mathbf{\overline{S}}_2 \right) \left(\bigotimes_{(x,2)(x,3)(x+\mu,0)(x+\mu,1)} \mathbf{\overline{T}}_\mu \right)\left( ~~\bigotimes_x \mathbf{\overline{S}}_1 \right) \left(\bigotimes_x \mathbf{\overline{B}} \right)\right] \\
    &  \left[\left(\bigotimes_x \mathbf{\overline{S}}_2~~ \right) \left(\bigotimes_{(x,2)(x,3)(x+\nu,0)(x+\nu,1)} \mathbf{\overline{T}}_\nu \right)\left( ~~\bigotimes_x \mathbf{\overline{S}}_1 \right) \left(\bigotimes_x \mathbf{\overline{B}} \right)\right] \\
    &  \left[\left(\bigotimes_x \mathbf{\overline{S}}_2~~ \right) \left(\bigotimes_{(x,2)(x,3)(x+\kappa,0)(x+\kappa,1)} \mathbf{\overline{T}}_\kappa \right)\left( ~~\bigotimes_x \mathbf{\overline{S}}_1 \right)\right] 
\label{eq:fullqca3d}
\end{align}
}

\subsection{Electric and magnetic contribution}
The electric contribution works in the same way as in the $2+1$ dimensional case, i.e. the exponentiated squared electric operator is applied at every link:
\begin{equation} \label{eq:electric3d}
    \mathbf{D_E} = \bigotimes_{x,\eta\in\{\mu,\nu,\kappa\}} e^{\frac{i}{2}\epsilon^2 g_E^2 E_{x:\eta}^2}
\end{equation} 

The magnetic contribution needs to be generalized to take into account the three dimensions, and thus the three possible directions for the plaquettes. Therefore, the magnetic contribution of the $3+1$ QED QCA is the same as in the $2+1$ case, but it is applied three times: one for each pair of directions. Let $\mathbf{D_{M,\eta,\zeta}}$ from Eq. \eqref{eq:qcamag1} or \eqref{eq:qcamag2} (depending on the formulation one chooses) denote the magnetic contribution along the two spatial dimension $\eta$ and $\zeta$. We now have:
\begin{equation}\label{eq:magnetic3d}
    \mathbf{\overline{D}_M} = \mathbf{D_{M,\mu,\nu}D_{M,\mu,\kappa}D_{M,\nu,\kappa}}.
\end{equation}

Again this evolution coincides, in the limit, with the magnetic part of the Kogut-Susskind Hamiltonian:
\begin{align}
    \mathcal{H}_m =& \frac{g_M^2}{2}\sum_{\substack{x \\ \eta,\zeta\in\{\mu, \nu,\kappa\}\\ \eta\neq\zeta }} P_{x:\eta,\zeta} \\
        =&\frac{g_M^2}{2}\sum_x \left(P_{x:\mu,\nu} + P^\dagger_{x:\mu,\nu}\right)
        +&\frac{g_M^2}{2}\sum_x \left(P_{x:\mu,\kappa} + P^\dagger_{x:\mu,\kappa}\right)
        +&\frac{g_M^2}{2}\sum_x \left(P_{x:\nu,\kappa} + P^\dagger_{x:\nu,\kappa}\right).
\end{align}

\subsection{Complete dynamics}
Combining the $3+1$ Dirac QCA of \eqref{eq:fullqca3d} with the electric \eqref{eq:electric3d} and magnetic \eqref{eq:magnetic3d} contributions, one obtains the $3+1$ QED QCA:
\begin{equation}\label{eq:qca3d}
    \overline{\text{\bf{QCA}}} = \mathbf{\overline{D}_M\overline{D}_E\overline{D}_F}.
\end{equation}

Its gauge invariance is ensured by the same arguments as in the $2+1$ case.

\section{Conclusion}
\paragraph{Summary of contributions.}
In this paper we constructed a quantum cellular automata (QCA) accounting for QED in $2+1$ and $3+1$ dimensions. The construction follows the same principles used to build the Lagrangian formulation of QED---i.e. free anti-commuting fermions, gauge invariant, simplest electric and magnetic term. But here spacetime is discrete, and space and time are treated on equal footing. The evolution is described in terms of local quantum gates, whose wirings coincide exactly with the speed of light of the QED. To reach our goal, we needed three contributions. 

The first contribution was the formulation of gauge invariant local evolution operators---Eqs. \eqref{eq:localhoppingterm} and \eqref{eq:gaugelocality}---that meet the specifications imposed by the (anti-)com\-mutation relations \eqref{eq:fermioncommutation}, \eqref{eq:bosoncommutation} and \eqref{eq:bosonfermioncommutation} of the fermionic and bosonic annihilators/lowering operators they are made of. It was, to us, a surprise that this could be achieved since it is in apparent contradiction with the no-go result of \cite{Mlodinow2020QuantumFT}. But, following ideas of \cite{farrelly2017insights,Magnifico2021LatticeQE} the gauge field came to the rescue. This was used to obtain a $2+1$ Dirac QCA, i.e. a generalization of the $2+1$ Dirac QW to multiple walkers, recovering the free fermionic dynamics $\mathbf{D_F}$ \eqref{eq:qcafermion} which is represented in Fig. \ref{fig:fermionicevolution}. 

The second contribution was to derive the electric $\mathbf{D_E}$ \eqref{eq:qcaelectric} and magnetic $\mathbf{D_M}$ \eqref{eq:qcamag1} contributions, leading to the $2+1$ QED QCA in Eq. \eqref{eq:fullqca2d}. The magnetic contributions was the most challenging, but in the end two possible quantum circuit representations were found. The first uses a diagonalization, through a Fourier transform, in order to exponentiate plaquette terms \eqref{eq:qcamag1first}, albeit up to truncation. The second approach formulates the magnetic contribution in terms of a quantum walk over the gauge field Hilbert space $\mathcal{H}_\mathbb{Z}$ \eqref{eq:qcamag2}. Both approaches lead to the same continuum limit \eqref{eq:qcamag1limit} and were checked to correspond to integration of the magnetic part of the Kogut-Susskind Hamiltonian. This provides a first---to the authors' knowledge---discrete spacetime formulation of the magnetic term.

The third contribution was the extension of the QCA to a $3+1$ QED QCA \eqref{eq:qca3d} which required raising the state space from $2$ to $4$ qubits per sites to represent fermions. On paper, this meant working out the consequences of the anti-commutation relations for up to four fermions, under sophisticated changes of basis. Fortunately, these $4$-qubit gates were decomposable as $2$-qubit gates. The electric and magnetic contribution were straightforwardly extended from the $2+1$ case. Altogether, the $3+1$ QED QCA provides a first, relativistic discrete spacetime formulation of a real-life quantum field theory.

\PA{It is not clear whether the proposed $2+1$ QED QCA and $3+1$ QED QCA hereby constructed are `staggered' in the traditional sense, because particles and antiparticles live at the same positions. But they are at least staggered-like in the sense of the $1+1$ QED QCA of \cite{Arrighi2020AQC}. In $1+1$, this suffices to escape the fermion-doubling problem, whereby high momentum wavepackets result in low energy states. We leave it as an open question whether fermion-doubling is escaped in a similar way in the proposed $2+1$ and $3+1$ models.}

\paragraph{Perspectives.}
These two and three-dimensional QED QCA are quantum circuits. $\mathbf{D_F}$, $\mathbf{D_E}$ and $\mathbf{D_M}$ can be expressed in terms of standard universal gates such as \textsc{CNot}, \textsc{Phase}, \textsc{Hadamard}. Thus, the QCA is directly interpretable as a digital quantum simulation algorithm, to be run on a Quantum Computer. A first perspective is the implementation of this QCA on quantum computers. This simulation scheme is efficient, in that it requires $O(s^d/\Delta_x^d\ t/\Delta_t)$ gates in order to simulate a chunk of space of size $s$, over $t$ time steps with $\Delta_x$ the space resolution, $\Delta_t$ the time resolution, and $d$ the space dimension. The output of this is a quantum state and the simulation may need to be run multiple times in order to obtain meaningful statistics. However, classically just the state space itself is an $O(e^{s^d/\Delta_x^d})$ as it grows exponentially with the number of quantum systems to be simulated. The classical time complexity is thus $O(e^{s^d/\Delta_x^d}t/\Delta_t)$. The exponential gain here clearly comes from the fact that the scheme simulates a multi-particle systems, just like in Hamiltonian-based multi-particle quantum simulation schemes. Quantum walk-based simulation schemes on the other hand, are by definition in the one-particle sector, and thus can only yield polynomial gains.

An immediate continuation of this work would be to further parametrize the QED QCA, so as to make it `plastic' enough so that we may be able to take a discrete space continuous time limit of the model, and prove that one recovers the Kogut-Susskind Hamiltonian, as was done for $1+1$-QED in \cite{Sellapillay2022ADR}. In this paper, the QFT considered is an abelian gauge theory. Another extension of this work is to consider non-abelian gauge transformations so as to recover the quantum chromodynamics (QCD). Looking at different geometries for the underlying space, another extension would be to define the QCA over a triangular or tetrahedral spatial discretization of space, as was done for quantum walks \cite{ArrighiTriang}. Finally, notice that the distinction between fermions and interacting-hardcore-bosons \cite{PerinottiFermions,MarlettoFermions} is wearing thin with this local model, and yet seems to persist as embodied by the $Z$ terms of Eqs. \eqref{eq:localhoppingterm} and \eqref{eq:cornerop}. This is intriguing, and one truly wonders whether the distinction is physically observable, or can be proven otherwise.

\paragraph{Acknowledgements.} We warmly thank Dogukan Bakircioglu for pointing out an earlier mistake at Eq. \eqref{eq:3Dtransport}, now patched in this v4. The authors would like to thank Pablo Arnault, Cédric Bény, Terry Farrelly and Simone Montangero for helpful conversations. This project/publication was made possible through the support of the ID\# 62312 grant from the John Templeton Foundation, as part of the \href{https://www.templeton.org/grant/the-quantum-information-structure-of-spacetime-qiss-second-phase}{‘The Quantum Information Structure of Spacetime’ Project (QISS) }. The opinions expressed in this project/publication are those of the author(s) and do not necessarily reflect the views of the John Templeton Foundation. It is also supported by the PEPR integrated project EPiQ ANR-22-PETQ-0007 part of Plan France 2030. G. M. is partially supported by the Italian PRIN2017, the Horizon 2020 research and innovation programme under grant agreement No 817482 (Quantum Flagship - PASQuanS), the INFN project QUANTUM, and by European Union - NextGenerationEU project CN00000013 - Italian Research Center on HPC, Big Data and Quantum Computing.

\bibliographystyle{quantum}
\bibliography{biblio}

\begin{thebibliography}{10}

\bibitem{lloyd1996universal}
Seth Lloyd.
\newblock ``Universal quantum simulators''.
\newblock \href{https://dx.doi.org/10.1126/science.273.5278.1073}{Science {\bf
  273}, 1073--1078}~(1996).

\bibitem{jordan2012quantum}
Stephen~P Jordan, Keith~SM Lee, and John Preskill.
\newblock ``Quantum algorithms for quantum field theories''.
\newblock \href{https://dx.doi.org/10.1126/science.1217069}{Science {\bf 336},
  1130--1133}~(2012).

\bibitem{banuls2020simulating}
Mari~Carmen Banuls, Rainer Blatt, Jacopo Catani, Alessio Celi, Juan~Ignacio
  Cirac, Marcello Dalmonte, Leonardo Fallani, Karl Jansen, Maciej Lewenstein,
  Simone Montangero, et~al.
\newblock ``Simulating lattice gauge theories within quantum technologies''.
\newblock \href{https://dx.doi.org/10.1140/epjd/e2020-100571-8}{The European
  physical journal D {\bf 74}, 1--42}~(2020).

\bibitem{Preskill2019SimulatingQF}
John Preskill.
\newblock ``Simulating quantum field theory with a quantum computer''.
\newblock \href{https://dx.doi.org/10.22323/1.334.0024}{Proceedings of The 36th
  Annual International Symposium on Lattice Field Theory —
  PoS(LATTICE2018)}~(2019).

\bibitem{Kitaev2003FaultTQ}
Alexei~Y. Kitaev.
\newblock ``Fault tolerant quantum computation by anyons''.
\newblock \href{https://dx.doi.org/10.1016/S0003-4916(02)00018-0}{Annals of
  Physics {\bf 303}, 2--30}~(2003).

\bibitem{Savary2017QuantumSL}
Lucile Savary and Leon Balents.
\newblock ``Quantum spin liquids: a review.''.
\newblock \href{https://dx.doi.org/10.1088/0034-4885/80/1/016502}{Reports on
  progress in physics. Physical Society {\bf 80 1}, 016502}~(2017).

\bibitem{Gonzalez2022hardware}
Daniel Gonz\'alez-Cuadra, Torsten~V. Zache, Jose Carrasco, Barbara Kraus, and
  Peter Zoller.
\newblock ``Hardware efficient quantum simulation of non-abelian gauge theories
  with qudits on rydberg platforms''.
\newblock \href{https://dx.doi.org/10.1103/PhysRevLett.129.160501}{Phys. Rev.
  Lett. {\bf 129}, 160501}~(2022).

\bibitem{knechtli2017lattice}
Francesco Knechtli, Michael G{\"u}nther, and Michael Peardon.
\newblock ``Lattice quantum chromodynamics: practical essentials''.
\newblock \href{https://dx.doi.org/10.1007/978-94-024-0999-4}{Springer}.
  ~(2017).

\bibitem{Kogut1975HamiltonianFO}
J.~B. Kogut and Leonard Susskind.
\newblock ``Hamiltonian formulation of wilson's lattice gauge theories''.
\newblock \href{https://dx.doi.org/10.1103/PhysRevD.11.395}{Physical Review D
  {\bf 11}, 395--408}~(1975).

\bibitem{Banks1976StrongCC}
Thomas Banks, Leonard Susskind, and J.~B. Kogut.
\newblock ``Strong coupling calculations of lattice gauge theories:
  (1+1)-dimensional exercises''.
\newblock \href{https://dx.doi.org/10.1103/PhysRevD.13.1043}{Physical Review D
  {\bf 13}, 1043--1053}~(1976).

\bibitem{Martinez2016RealtimeDO}
Esteban~A. Martinez, Christine~A. Muschik, Philipp Schindler, Daniel Nigg,
  Alexander Erhard, Markus Heyl, Philipp Hauke, Marcello Dalmonte, Thomas Monz,
  Peter Zoller, and Rainer Blatt.
\newblock ``Real-time dynamics of lattice gauge theories with a few-qubit
  quantum computer''.
\newblock \href{https://dx.doi.org/10.1038/nature18318}{Nature {\bf 534},
  516--519}~(2016).

\bibitem{Magnifico2021LatticeQE}
Giuseppe Magnifico, Timo Felser, Pietro Silvi, and Simone Montangero.
\newblock ``Lattice quantum electrodynamics in (3+1)-dimensions at finite
  density with tensor networks''.
\newblock \href{https://dx.doi.org/10.1038/s41467-021-23646-3}{Nature
  Communications{\bf 12}}~(2021).

\bibitem{Ors2013API}
Rom{\'a}n Or{\'u}s.
\newblock ``A practical introduction to tensor networks: Matrix product states
  and projected entangled pair states''.
\newblock \href{https://dx.doi.org/10.1016/j.aop.2014.06.013}{Annals of Physics
  {\bf 349}, 117--158}~(2013).

\bibitem{Byrnes2002DensityMR}
Tim Byrnes, Pranav Sriganesh, Robert~J. Bursill, and Chris~J. Hamer.
\newblock ``Density matrix renormalization group approach to the massive
  schwinger model''.
\newblock \href{https://dx.doi.org/10.1103/PhysRevD.66.013002}{Physical Review
  D {\bf 66}, 013002}~(2002).

\bibitem{Zapp2017TensorNS}
Kai Zapp and Rom{\'a}n Or{\'u}s.
\newblock ``Tensor network simulation of qed on infinite lattices: Learning
  from (1+1)d , and prospects for (2+1)d''.
\newblock \href{https://dx.doi.org/10.1103/PhysRevD.95.114508}{Physical Review
  D{\bf 95}}~(2017).

\bibitem{Arrighi2014TheDE}
Pablo Arrighi, M.~Forets, and Vincent Nesme.
\newblock ``The dirac equation as a quantum walk: higher dimensions,
  observational convergence''.
\newblock \href{https://dx.doi.org/10.1088/1751-8113/47/46/465302}{Journal of
  Physics A {\bf 47}, 465302}~(2014).

\bibitem{arrighi2014discrete}
P.~Arrighi, Stefano Facchini, and Marcelo Forets.
\newblock ``Discrete lorentz covariance for quantum walks and quantum cellular
  automata''.
\newblock \href{https://dx.doi.org/10.1088/1367-2630/16/9/093007}{New Journal
  of Physics {\bf 16}, 093007}~(2014).

\bibitem{PaviaLORENTZ2}
Alessandro Bisio, Giacomo~Mauro D’Ariano, and Paolo Perinotti.
\newblock ``Quantum walks, weyl equation and the lorentz group''.
\newblock \href{https://dx.doi.org/10.1007/s10701-017-0086-3}{Foundations of
  Physics {\bf 47}, 1065--1076}~(2017).

\bibitem{DebbaschLORENTZ}
Fabrice Debbasch.
\newblock ``Action principles for quantum automata and lorentz invariance of
  discrete time quantum walks''.
\newblock \href{https://dx.doi.org/10.1016/j.aop.2019.03.005}{Annals of Physics
  {\bf 405}, 340--364}~(2019).

\bibitem{osborne2019continuum}
Tobias~J Osborne.
\newblock ``Continuum limits of quantum lattice systems''~(2019).

\bibitem{eisert2009supersonic}
Jens Eisert and David Gross.
\newblock ``Supersonic quantum communication''.
\newblock \href{https://dx.doi.org/10.1103/PhysRevLett.102.240501}{Physical
  review letters {\bf 102}, 240501}~(2009).

\bibitem{cheneau2020speed}
Marc Cheneau and Laurent Sanchez-Palencia.
\newblock ``A speed test for ripples in a quantum system''.
\newblock
  \href{https://dx.doi.org/http://dx.doi.org/10.1103/Physics.13.109}{Physics
  {\bf 13}, 109}~(2020).

\bibitem{Schwinger1962GaugeIA}
Julian~Seymour Schwinger.
\newblock ``Gauge invariance and mass''.
\newblock \href{https://dx.doi.org/10.1103/PhysRev.125.397}{Physical Review
  {\bf 125}, 2425--2429}~(1962).

\bibitem{Arrighi2020AQC}
Pablo Arrighi, C{\'e}dric B{\'e}ny, and Terry Farrelly.
\newblock ``A quantum cellular automaton for one-dimensional qed''.
\newblock \href{https://dx.doi.org/10.1007/s11128-019-2555-4}{Quantum
  Information Processing {\bf 19}, 1--28}~(2020).

\bibitem{di2020quantum}
Giuseppe Di~Molfetta and Pablo Arrighi.
\newblock ``A quantum walk with both a continuous-time limit and a
  continuous-spacetime limit''.
\newblock \href{https://dx.doi.org/10.1007/s11128-019-2549-2}{Quantum
  Information Processing {\bf 19}, 1--16}~(2020).

\bibitem{manighalam2021continuous}
Michael Manighalam and Giuseppe Di~Molfetta.
\newblock ``Continuous time limit of the dtqw in 2d+ 1 and plasticity''.
\newblock \href{https://dx.doi.org/10.1007/s11128-021-03011-5}{Quantum
  Information Processing {\bf 20}, 1--24}~(2021).

\bibitem{Sellapillay2022ADR}
Kevissen Sellapillay, Pablo Arrighi, and Giuseppe~Di Molfetta.
\newblock ``A discrete relativistic spacetime formalism for 1 + 1-qed with
  continuum limits''.
\newblock \href{https://dx.doi.org/10.1038/s41598-022-06241-4}{Scientific
  Reports{\bf 12}}~(2022).

\bibitem{di2013quantum}
Giuseppe Di~Molfetta, Marc Brachet, and Fabrice Debbasch.
\newblock ``Quantum walks as massless dirac fermions in curved space-time''.
\newblock \href{https://dx.doi.org/10.1103/PhysRevA.88.042301}{Physical Review
  A {\bf 88}, 042301}~(2013).

\bibitem{di2016quantum}
Giuseppe Di~Molfetta and Armando P{\'e}rez.
\newblock ``Quantum walks as simulators of neutrino oscillations in a vacuum
  and matter''.
\newblock \href{https://dx.doi.org/10.1088/1367-2630/18/10/103038}{New Journal
  of Physics {\bf 18}, 103038}~(2016).

\bibitem{hatifi2019quantum}
Mohamed Hatifi, Giuseppe Di~Molfetta, Fabrice Debbasch, and Marc Brachet.
\newblock ``Quantum walk hydrodynamics''.
\newblock \href{https://dx.doi.org/10.1038/s41598-019-40059-x}{Scientific
  reports {\bf 9}, 1--7}~(2019).

\bibitem{ahlbrecht2012molecular}
Andre Ahlbrecht, Andrea Alberti, Dieter Meschede, Volkher~B Scholz, Albert~H
  Werner, and Reinhard~F Werner.
\newblock ``Molecular binding in interacting quantum walks''.
\newblock \href{https://dx.doi.org/10.1088/1367-2630/14/7/073050}{New Journal
  of Physics {\bf 14}, 073050}~(2012).

\bibitem{PaviaMolecular}
Alessandro Bisio, Giacomo~Mauro D'Ariano, Paolo Perinotti, and Alessandro
  Tosini.
\newblock ``Thirring quantum cellular automaton''.
\newblock \href{https://dx.doi.org/10.1103/PhysRevA.97.032132}{Physical Review
  A {\bf 97}, 032132}~(2018).

\bibitem{Mlodinow2020QuantumFT}
Leonard~D. Mlodinow and Todd~A. Brun.
\newblock ``Quantum field theory from a quantum cellular automaton in one
  spatial dimension and a no-go theorem in higher dimensions''.
\newblock \href{https://dx.doi.org/10.1103/PhysRevA.102.042211}{Physical Review
  A}~(2020).

\bibitem{brun2020quantum}
Todd~A Brun and Leonard Mlodinow.
\newblock ``Quantum cellular automata and quantum field theory in two spatial
  dimensions''.
\newblock \href{https://dx.doi.org/10.1103/PhysRevA.102.062222}{Physical Review
  A {\bf 102}, 062222}~(2020).

\bibitem{mlodinow2021fermionic}
Leonard Mlodinow and Todd~A Brun.
\newblock ``Fermionic and bosonic quantum field theories from quantum cellular
  automata in three spatial dimensions''.
\newblock \href{https://dx.doi.org/10.1103/PhysRevA.103.052203}{Physical Review
  A {\bf 103}, 052203}~(2021).

\bibitem{Zohar2018EliminatingFM}
Erez Zohar and Juan~Ignacio Cirac.
\newblock ``Eliminating fermionic matter fields in lattice gauge theories''.
\newblock \href{https://dx.doi.org/10.1103/PhysRevB.98.075119}{Physical Review
  B}~(2018).

\bibitem{farrelly2017insights}
Terry Farrelly.
\newblock ``Insights from quantum information into fundamental
  physics''~(2017).

\bibitem{arrighi2018gauge}
Pablo Arrighi, Giuseppe~Di Molfetta, and Nathana{\"e}l Eon.
\newblock ``A gauge invariant reversible cellular automaton''.
\newblock In International Workshop on Cellular Automata and Discrete Complex
  Systems.
\newblock \href{https://dx.doi.org/10.1007/978-3-319-92675-9_1}{Pages 1--12}.
\newblock Springer~(2018).

\bibitem{Arrighi2022GaugeinvarianceIC}
Pablo Arrighi, Giuseppe~Di Molfetta, and Nathanael Eon.
\newblock ``Gauge invariance in cellular automata''.
\newblock \href{https://dx.doi.org/10.1007/s11047-022-09879-1}{Natural
  ComputingPages 1--13}~(2022).

\bibitem{ArrighiGaugeUniversality}
Pablo Arrighi, Marin Costes, and Nathana\"{e}l Eon.
\newblock ``{Universal Gauge invariant Cellular Automata}''.
\newblock In Filippo Bonchi and Simon~J. Puglisi, editors, 46th International
  Symposium on Mathematical Foundations of Computer Science (MFCS 2021).
\newblock \href{https://dx.doi.org/10.4230/LIPIcs.MFCS.2021.9}{Volume 202 of
  Leibniz International Proceedings in Informatics (LIPIcs), pages 9:1--9:14}.
\newblock Dagstuhl, Germany~(2021). Schloss Dagstuhl -- Leibniz-Zentrum f{\"u}r
  Informatik.

\bibitem{MolfettaGaugeQW}
Giuseppe Di~Molfetta, Marc Brachet, and Fabrice Debbasch.
\newblock ``Quantum walks in artificial electric and gravitational fields''.
\newblock \href{https://dx.doi.org/10.1016/j.physa.2013.11.036}{Physica A:
  Statistical Mechanics and its Applications {\bf 397}, 157--168}~(2014).

\bibitem{Melnikov2000LatticeSM}
Kirill Melnikov and Marvin Weinstein.
\newblock ``Lattice schwinger model: Confinement, anomalies, chiral fermions,
  and all that''.
\newblock \href{https://dx.doi.org/10.1103/PhysRevD.62.094504}{Physical Review
  D {\bf 62}, 094504}~(2000).

\bibitem{Magnifico2020realtimedynamics}
Giuseppe Magnifico, Marcello Dalmonte, Paolo Facchi, Saverio Pascazio,
  Francesco~V. Pepe, and Elisa Ercolessi.
\newblock ``Real {T}ime {D}ynamics and {C}onfinement in the {$\mathbb{Z}_{n}$}
  {S}chwinger-{W}eyl lattice model for 1+1 {QED}''.
\newblock \href{https://dx.doi.org/10.22331/q-2020-06-15-281}{{Quantum} {\bf
  4}, 281}~(2020).

\bibitem{di2012discrete}
Giuseppe Di~Molfetta and Fabrice Debbasch.
\newblock ``Discrete-time quantum walks: Continuous limit and symmetries''.
\newblock \href{https://dx.doi.org/10.1063/1.4764876}{Journal of Mathematical
  Physics {\bf 53}, 123302}~(2012).

\bibitem{ercolessi2018phase}
Elisa Ercolessi, Paolo Facchi, Giuseppe Magnifico, Saverio Pascazio, and
  Francesco~V Pepe.
\newblock ``Phase transitions in z n gauge models: towards quantum simulations
  of the schwinger-weyl qed''.
\newblock \href{https://dx.doi.org/10.1103/PhysRevD.98.074503}{Physical Review
  D {\bf 98}, 074503}~(2018).

\bibitem{Haase2021ARE}
Jan~F. Haase, Luca Dellantonio, Alessio Celi, Danny Paulson, Angus Kan, Karl
  Jansen, and Christine~A. Muschik.
\newblock ``A resource efficient approach for quantum and classical simulations
  of gauge theories in particle physics''.
\newblock \href{https://dx.doi.org/10.22331/q-2021-02-04-393}{Quantum {\bf 5},
  393}~(2021).

\bibitem{ShorQFT}
Peter~W Shor.
\newblock ``Polynomial-time algorithms for prime factorization and discrete
  logarithms on a quantum computer''.
\newblock \href{https://dx.doi.org/10.1137/S0036144598347011}{SIAM review {\bf
  41}, 303--332}~(1999).

\bibitem{marquez2017fermion}
Ivan M{\'a}rquez-Mart{\'\i}n, Giuseppe Di~Molfetta, and Armando P{\'e}rez.
\newblock ``Fermion confinement via quantum walks in (2+ 1)-dimensional and (3+
  1)-dimensional space-time''.
\newblock \href{https://dx.doi.org/10.1103/PhysRevA.95.042112}{Physical Review
  A {\bf 95}, 042112}~(2017).

\bibitem{ArrighiTetrahedra}
Ugo Nzongani, Nathana\"el Eon, Iv\'an M\'arquez-Mart\'{\i}n, Armando P\'erez,
  Giuseppe Di~Molfetta, and Pablo Arrighi.
\newblock ``Dirac quantum walk on tetrahedra''.
\newblock \href{https://dx.doi.org/10.1103/PhysRevA.110.042418}{Phys. Rev. A
  {\bf 110}, 042418}~(2024).

\bibitem{ArrighiTriang}
Pablo Arrighi, Giuseppe {Di\ Molfetta}, Iv{\'a}n M{\'a}rquez, and Armando
  P{\'e}rez.
\newblock ``Dirac equation as a quantum walk over the honeycomb and triangular
  lattices''.
\newblock \href{https://dx.doi.org/10.1103/PhysRevA.97.062111}{Physical Review
  A {\bf 97}, 062111}~(2018).

\bibitem{PerinottiFermions}
Matteo Lugli, Paolo Perinotti, and Alessandro Tosini.
\newblock ``Fermionic state discrimination by local operations and classical
  communication''.
\newblock \href{https://dx.doi.org/10.1103/PhysRevLett.125.110403}{Physical
  Review Letters {\bf 125}, 110403}~(2020).

\bibitem{MarlettoFermions}
Chiara Marletto and Vlatko Vedral.
\newblock ``Spin, statistics, spacetime and quantum gravity''~(2021).

\end{thebibliography}

\appendix
\section{From quantum walk to quantum cellular automata operators}
\label{appendix:qwtoqca}

We explain the process of extending a one-particle sector on-site operator making up a quantum walk, into a multi-particle sector on-site operator making up a quantum cellular automata. We start with the $2+1$ dimensional case.

The QCA on-site operators act on $2$ qubits and are number conserving, which constrains them. Indeed, the evolution for the input $\ket{00}^{(x,1)(x,0)}$ will always be the identity without loss of generality, and the evolution for the input $\ket{11}^{(x,1)(x,0)}$ is only a phase since it is the only state with occupation number equal to $2$. Therefore, any QCA on-site operator $\mathbf{W}$ can be written as a direct sum $\mathbf{W}=W \oplus e^{i\varphi}$ with $W=1\oplus M$ the quantum walk on-site operator for the one-particle sector and $\varphi$ a phase to be determined. To find out this phase exactly, we use the Heisenberg picture.

Let $\mathbf{W}=1\oplus M \oplus e^{i\varphi}$ be a QCA on-site operator as acting over some $\ket{\psi}^{(x,0),(x,1)}$ with:
\begin{equation}
    M = \begin{pmatrix}
        M_{00} & M_{01} \\
        M_{10} & M_{11}
    \end{pmatrix}
    = \begin{pmatrix}
        M_{00} & -e^{i\theta} M_{10}^* \\
        M_{10} &  e^{i\theta} M_{00}^*
    \end{pmatrix} ~~\textrm{without loss of generality.}
\end{equation}
The Heisenberg picture describes the future impact of our past actions. Consider the past action $a^\dagger_{x,1}$ at $t$:
\begin{align}
    a_{x,1}^\dagger &= \ket{1}^{x,1}\bra{0} \otimes Z_{x,0} \bigotimes_{y\prec (x,0)} Z_y\\
        &= (\ket{10}\bra{00} - \ket{11}\bra{01}) \bigotimes_{y\prec (x,0)} Z_y.
\end{align}
Its future impact at time $t+1$ is
\begin{align}
    \mathbf{W} a^\dagger_{x,1} \mathbf{W}^\dagger &=  \left[\left(M_{11}\ket{10} + M_{01}\ket{01}\right) \bra{00}- e^{i\varphi} \ket{11}\left(\bra{01}M_{00}^*+\bra{10}M_{10}^*) \right) \right]\bigotimes_{y\prec (x,0)} Z_y \\
    &=\left[ \ket{1}^{x,1}\bra{0}\otimes\begin{pmatrix}M_{11} & 0 \\ 0 & -e^{i\varphi} M_{00}^* \end{pmatrix} + \begin{pmatrix}M_{01} & 0 \\ 0 & -e^{i\varphi} M_{10}^*  \end{pmatrix}\otimes\ket{1}^{x,0}\bra{0}\right] \bigotimes_{y\prec (x,0)} Z_y.\\
    &=\left[ \ket{1}^{x,1}\bra{0}\otimes\begin{pmatrix}e^{i\theta} M_{00}^* & 0 \\ 0 & -e^{i\varphi} M_{00}^* \end{pmatrix} + \begin{pmatrix}-e^{i\theta} M_{10}^* & 0 \\ 0 & -e^{i\varphi} M_{10}^*  \end{pmatrix}\otimes\ket{1}^{x,0}\bra{0}\right] \bigotimes_{y\prec (x,0)} Z_y.
\end{align}
Suppose that the $\varphi$ we seek to determine, is equal to $\theta$. With this supposition, the above simplifies and we have that the future impact of $a^\dagger_{x,1}$ is just $\mathbf{W} a^\dagger_{x,1} \mathbf{W}^\dagger = M_{11} a_{1}^\dagger + M_{01} a_{x,0}^\dagger$.\\
A contrario, with any other choice of $\varphi$ we would be constructing some $\mathbf{W'}$ which, albeit coinciding with $\mathbf{W}$ and thus $W$ in the one-particle sector, would for instance leave a phase between $a_{x,1}^\dagger a_{x,0}^\dagger a_{x,0}$ and $a_{x,1}^\dagger a_{x,0} a_{x,0}^\dagger$ in the future impact of $a_{x,1}^\dagger$. The future impact of $a_{x,1}^\dagger$ would again yield a superposition of a particle at $(x,0)$ or at $(x,1)$, but with a phase depending on another particle being there or not. Such a $\mathbf{W'}$ would not, therefore, be the `rightful non-interacting extensions of $W$' to the multi-particle sector. Only $\mathbf{W}$ is. In other words, we were seeking to determine the $\varphi$ of the last entry of $\mathbf{W}$; setting it to $\theta$ fixes it to its non-interactive value.  

The same process can be applied to the evolution $a_{x,0}^\dagger$ leading to
\begin{align}
    \mathbf{W}a_{x,0}^\dagger\mathbf{W}^\dagger &= M_{00} a_{x,0}^\dagger + M_{10}a_{x,1}^\dagger\\
    \mathbf{W}a_{x,1}^\dagger\mathbf{W}^\dagger &= M_{01} a_{x,0}^\dagger + M_{11}a_{x,1}^\dagger.
\end{align}

Considering as past action the product of these two operators $a_{x,0}^\dagger a_{x,1}^\dagger$. Its future impact is
\begin{align}
    \mathbf{W}a_{x,0}^\dagger\mathbf{W}^\dagger \mathbf{W} a_{x,1}^\dagger \mathbf{W}^\dagger   &= (M_{00} a_{x,0}^\dagger + M_{10} a_{x,1}^\dagger)(M_{01} a_{x,0}^\dagger + M_{11} a_{x,1}^\dagger) \\
        &= M_{00}M_{11} a_{x,0}^\dagger a_{x,1}^\dagger + M_{01}M_{10} a_{x,1}^\dagger a_{x,0}^\dagger \\
        &= (M_{00}M_{11} - M_{01}M_{10}) a_{x,0}^\dagger a_{x,1}^\dagger.
\end{align}

Now, because state $\ket{00}^{(x,1)(x,0)}$ evolves into $\ket{00}^{(x,1)(x,0)}$, and because $a^\dagger_{x,0}a^\dagger_{x,1}$ changes $\ket{00}^{(x,1)(x,0)}$ into $\ket{11}^{(x,1)(x,0)}$, it must be that $\ket{11}^{(x,1)(x,0)}$  evolves into  
$(M_{00}M_{11} - M_{01}M_{10}) \ket{11}^{(x,1)(x,0)}$. Hence, the phase $e^{i\varphi}=e^{i\theta}$ applied to $\ket{11}^{(x,1)(x,0)}$ can simply be written $M_{00}M_{11} - M_{01}M_{10}$. 

In other words we have that for any QW on-site operator $W=1\oplus M$ acting on qubits, the corresponding non-interactive multi-particle extension QCA on-site operator is
\begin{equation}\label{eq:woperatorqcafromqw}
    \mathbf{W} = \begin{pmatrix}
        1 & 0 & 0 & 0 \\
        0 & M_{00} & M_{01} & 0 \\
        0 & M_{10} & M_{11} & 0 \\
        0 & 0 & 0 & M_{00}M_{11} - M_{01}M_{10}
    \end{pmatrix}.
\end{equation}
Let us now use this to define the on-site and transport gates of the QCA.

\subsection{On-site operators}

First consider $\mathbf{S}=1\oplus X \oplus e^{i\varphi}$ with $X_{01}=X_{10}=1$ and $X_{00}=X_{11}=0$. We justified in Subsec. \ref{sec:fermionicdyn} that $\varphi$ needs be $\pi$. With the above this readily follows from $X_{00}X_{11} - X_{01}X_{10}=-1$.
Therefore $\mathbf{S}$ \PA{as acting over some $\ket{\psi}^{(x,0),(x,1)}$} is:
\begin{equation}
    \mathbf{S} = \begin{pmatrix}
        1 & 0 & 0 & 0 \\
        0 & 0 & 1 & 0 \\
        0 & 1 & 0 & 0 \\
        0 & 0 & 0 & -1 \\
    \end{pmatrix}.
\end{equation}

On-site QW gates that are mere permutations of qubits in the one-particle sector, can then be extended to the multi-particle sector directly by means of products of $\mathbf{S}$, in the same way that any permutation can be obtained from local transpositions.

For the mass term and the basis changes, this is not the case, but Eq. \eqref{eq:woperatorqcafromqw} again readily applies. As an example, the mass term $C_\epsilon=1\oplus C$ has $C_{00}C_{11} - C_{01}C_{10} = c^2 +s^2=1$,
which results in the following mass on-site operator for the QCA, \PA{expressed as acting over some $\ket{\psi}^{(x,0),(x,1)}$}:
\begin{equation}
    \mathbf{C_\epsilon} = \begin{pmatrix}
        1 & 0 & 0 & 0 \\
        0 & c & -s & 0 \\
        0 & s & c & 0 \\
        0 & 0 & 0 & 1
    \end{pmatrix}.
\end{equation}

\subsection{Transport operators}

In this paper the transport is implemented in two steps: first swap the qubits on-site so that the right-moving qubit is on the right-hand side of the site ($\mathbf{S}$), then hop qubits across adjacent sites ($\mathbf{T}_\eta$) whilst changing the gauge field accordingly.\\
In the one particle sector, the across-two-sites operator $T_\eta$, \PA{as acting on some $\ket{\psi}^{(x+\eta,0)(x,1)}$}, is given by:
\begin{equation}
    T_{\eta} = 1\oplus T \textrm{ with }T=\begin{pmatrix}
        0 & K_\eta \\
        K^\dagger_\eta & 0 \\
    \end{pmatrix}
\end{equation}
with 
\begin{equation}
    K_{\eta}=
    \left(\bigotimes_{y\in \llbracket x, (x,\eta) \llbracket } Z_y \right) 
    U_{x:\eta}
    \left(\bigotimes_{y\in \llbracket  x+\eta, (x+\eta, -\eta) \llbracket } Z_y \right)
\end{equation}
and $U_{x:\eta}$ the action of the gauge field lowering operator on $(x:\eta)(x+\eta:-\eta)$. We see that $T$ plays the same role as $M$ in Eq. \eqref{eq:woperatorqcafromqw}. The same reasoning applies. 
Since 
\begin{align}
    T_{00} T_{11} - T_{01} T_{10} = -K_\eta K^\dagger_\eta = -I
\end{align}
we have, \PA{as acting on some $\ket{\psi}^{(x+\eta,0)(x,1)}$}:
\begin{equation}
    \mathbf{T}_{\eta} = \begin{pmatrix}
        1 & 0 & 0 & 0 \\
        0 & 0 & K_{\eta} & 0 \\
        0 & K^\dagger_{\eta} & 0 & 0 \\
        0 & 0 & 0 & -1 \\
    \end{pmatrix}.
\end{equation}

\subsection{3+1 dimensions}
In $3+1$ dimensions, there are four qubits per site instead of two. A QW operators $\overline{W}$ that acted in the one particle sector $\ket{0001}^{(x,3)(x,2)(x,1)(x,0)}, \ket{0010}^{(x,3)(x,2)(x,1)(x,0)},\ldots$, now needs to be extended to a QCA operator $\mathbf{\overline{W}}$ that can handle the multi-particle sector $\ket{0011}^{(x,3)(x,2)(x,1)(x,0)}, \ket{1110}^{(x,3)(x,2)(x,1)(x,0)},\ldots$, which seems to leave open many more possibilities than in the $2+1$ case. However, the $3+1$ QW operators we are dealing with can easily be decomposed as circuits of $2$-qubit gates, with the qubits following each other in the JW order. So, we extend these two qubit gates instead, through the same process as in $2+1$ dimensions, and then recombine them to form the extension of the $3+1$ QW operator.

The $2$-qubit QW operators used will turn out to be same as those of the $2+1$ case, except for $1\oplus B$ which was not defined previously. Its QCA version $\mathbf{{B}}$ is given through Eq. \eqref{eq:woperatorqcafromqw}, yielding, \PA{as acting on some $\ket{\psi}^{(x,1)(x,0)}$}, Eq. \eqref{eq:2DqcaB}.

Let us first decompose the $4$-qubit operators for the $3+1$ QW that was given in Eq. \eqref{eq:3dqw}, as circuits of $2$-qubit gates.

\begin{align}
    \overline{C}_\epsilon &= 1\oplus(1\oplus X \oplus 1)(C_\epsilon\oplus C_\epsilon)(1\oplus X \oplus 1) \\ 
    \overline{S} &= 1\oplus(I\oplus X \oplus I)(X \oplus X)(I\oplus X \oplus I)\\
    \overline{T}_{\eta,\epsilon} &=1\oplus (I\oplus T_{\eta} \oplus I)(X \oplus X)(I\oplus T_{\eta} \oplus I)\\
\overline{B}&=1\oplus(B\oplus B)
\end{align}
These operators are written as a product of direct sums of gates. The direct sum is used for spatial composition and not the tensor, because we are in the one-particle sector. \PA{The matrices are expressed as acting upon $\ket{\psi}^{(x,3)(x,2)(x,1)(x,0)}$, i.e. w.r.t the canonical basis 
\begin{align*}
\{&\ket{0000}^{(x,3)(x,2)(x,1)(x,0)}, \ket{0001}^{(x,3)(x,2)(x,1)(x,0)}, \ket{0010}^{(x,3)(x,2)(x,1)(x,0)},\\ &\ket{0100}^{(x,3)(x,2)(x,1)(x,0)}, \ket{1000}^{(x,3)(x,2)(x,1)(x,0)}\}
\end{align*} 
in this order, and with $\ket{0000}^{(x,3)(x,2)(x,1)(x,0)}$ acted upon trivially.} Notice that each on-site operator on the right-hand-side acts on neighbouring qubits, e.g. in $\overline{C}_\epsilon$ the rightmost $C_\epsilon$ acts on $(x,1)(x,0)$ etc.
Indeed, whenever some $4$ qubit operator needed to act on non-adjacent qubits in the JW order, a prior reordering using $X$ was introduced. For instance, the mass term on-site operator, \PA{as acting on some $\ket{\psi}^{(x,3)(x,2)(x,1)(x,0)}$}, is given by:
$$\overline{C}_\epsilon = 1 \oplus \begin{pmatrix}
    c & 0 & -s & 0 \\
    0 & c & 0 & -s \\
    s & 0 & c & 0 \\
    0 & s & 0 & c \\
\end{pmatrix}.$$ 
One notices that it acts on the zeroth and second qubits on the one hand, and on the first and third qubits on the other. In order to obtain a circuit of adjacent gates in the JW order, The first and second qubits are swapped, using the operation $1\oplus X \oplus 1$, allowing for the application of the $2$-qubit mass term on the leading and last two qubits separately through $C_\epsilon \oplus C_\epsilon$. The swap is then reapplied such that the initial order is recovered. These swaps can be understood as crossing of wires in a circuit so that the $2$-qubit gates have the correct inputs and outputs. Such swaps correspond, in the multi-particle sector, to the QCA operator $\mathbf{S}$. Hence, we can extend  $1\oplus X \oplus 1$ into  $1\otimes \mathbf{S} \otimes 1$, as this again swaps the two middle qubits while leaving the rest unchanged. Notice that the operator $\otimes$ is used instead of $\oplus$ since we are no longer in the one particle sector. Similarly, the second step of the circuit for the mass term is extended into $\mathbf{C_\epsilon}\otimes \mathbf{C_\epsilon}$. In the end, one obtains the following multi-particle sector, QCA operator for the mass term, \PA{as acting on some $\ket{\psi}^{(x,3)(x,2)(x,1)(x,0)}$, (i.e. w.r.t the canonical basis 
\begin{align*}
\{&\ket{0000}^{(x,3)(x,2)(x,1)(x,0)}, \ket{0001}^{(x,3)(x,2)(x,1)(x,0)},\\ 
&\ket{0010}^{(x,3)(x,2)(x,1)(x,0)}, \ket{0011}^{(x,3)(x,2)(x,1)(x,0)},\ldots\}
\end{align*} in this order):}
\begin{equation}
    \mathbf{\overline{C}_\epsilon} = \left( 1\otimes \mathbf{S} \otimes 1 \right) \left( \mathbf{C_\epsilon}\otimes \mathbf{C_\epsilon} \right) \left( 1\otimes \mathbf{S} \otimes 1 \right).
\end{equation}
The same procedure readily yields the gates that compose the $3+1$ QED QCA of Sec. \ref{sec:3dfermdyn}.

\end{document}